\documentclass[a4paper,12pt]{article}
\usepackage[utf8]{inputenc}

\usepackage{amsmath,amsfonts,amsthm,amssymb,dsfont}
\usepackage{graphicx,wrapfig,lipsum}
\usepackage[section]{placeins}

\usepackage{comment}

\usepackage[numbers,sort&compress]{natbib}

\usepackage{tikz}
\usetikzlibrary{shapes.geometric,decorations.markings}
\usepackage{subfigure}

\usepackage[hidelinks]{hyperref}

\hypersetup{
    colorlinks=true,
    linktocpage=true,
    linkcolor=blue,
    urlcolor=blue,
    citecolor=blue,
}

\numberwithin{equation}{section}
\newcommand{\Sp}[1]{\text{S}^{#1}}

\newcommand{\Vol}{\text{Vol}}
\newcommand{\vol}{\text{vol}}
\newcommand{\core}{\text{core}}
\newcommand{\hol}{\text{hol}}
\newcommand{\renorm}{\text{renorm}}
\newcommand{\LUV}{\Lambda_{\text{UV}}}

\newcommand{\calR}{\mathcal{R}}
\newcommand{\AdS}{\text{AdS}}
\newcommand{\YM}{\text{YM}}
\newcommand{\dual}{\text{dual}}
\newcommand{\tDelta}{\tilde{\Delta}}
\newcommand{\calF}{\mathcal{F}}
\newcommand{\calA}{\mathcal{A}}
\newcommand{\calD}{\mathcal{D}}
\newcommand{\calCm}[1]{\mathcal{C}_{#1}}
\newcommand{\Gn}[1]{G_{N}^{(#1)}}
\newcommand{\floor}[1]{\left\lfloor #1 \right\rfloor}
\newcommand{\dd}{\mathrm{d}}
\newcommand{\RR}{\text{R}}
\newcommand{\FF}{\text{F}}

\begin{document}

\begin{titlepage}

\begin{center}

$\phantom{.}$\\ \vspace{2cm}
\noindent{\Large{\textbf{Monodromy Defects in Maximally Supersymmetric Yang-Mills Theories from Holography}}}

\vspace{1cm}

Andrea Conti \footnote{contiandrea@uniovi.es}, 
Ricardo Stuardo \footnote{ricardostuardotroncoso@gmail.com}

\vspace{0.5cm}

Departamento de Física, Universidad de Oviedo,
C/ Leopoldo Calvo Sotelo, 18, 33007 Oviedo, Spain \\[2mm]
and \\[2mm]
Instituto Universitario de Ciencias y Tecnologías Espaciales de Asturias (ICTEA), Calle de la Independencia 13, 33004 Oviedo, Spain \\[2mm]

\end{center}

\vspace{0.5cm}
\centerline{\textbf{Abstract}} 

\vspace{0.5cm}

\noindent We study three Type II supergravity solutions holographically dual to codimension-2 supersymmetric defects in $(p+1)$-dimensional SU($N$) maximally supersymmetric Yang-Mills theory ($p=2,3,4$). In all of these cases, the defects have a non-trivial monodromy for the maximal abelian subgroup of the SO($9-p$) R-symmetry. Such solutions are obtained by considering branes wrapping spindle configurations, changing the parameters (which alters the coordinate domain), and imposing suitable boundary conditions. We provide a prescription to compute the entanglement entropy of the effective theory on the defect. We find the resulting quantity to be proportional to the free energy of the ambient theory. A similar analysis is performed for the D5-brane wrapping a spindle, but we find that changing the coordinate domain does not lead to a defect solution, but rather to a circle compactification.

\vspace*{\fill}

\end{titlepage}

\newpage

\tableofcontents
\thispagestyle{empty}

\newpage
\setcounter{page}{1}
\setcounter{footnote}{0}

\section{Introduction}

The AdS/CFT correspondence \cite{Maldacena:1997re,Gubser:1998bc,Witten:1998qj} and its generalization to non-conformal theories \cite{Itzhaki:1998dd,Boonstra:1998mp} have become a fruitful framework to study different aspects of supersymmetric gauge theories in various dimensions. Information about the dynamics of the gauge theory, operator spectrum and observables is geometrically encoded in the dual supergravity background, which allows for an exact or semi-analytic computation of these quantities. 

Recently, there has been an increasing interest in the study of defects in quantum field theories. See for example \cite{Billo:2016cpy,Wang:2020xkc,Chalabi:2021jud,Rodriguez-Gomez:2022gif,Rodriguez-Gomez:2022gbz,Bianchi:2015liz,Bianchi:2019umv,Bianchi:2019sxz,Bianchi:2021snj,CarrenoBolla:2023sos,CarrenoBolla:2023vrv,Mauch:2025irx,Calvo:2025kjh,Giombi:2021uae} and references therein. Of particular interest are defects in supersymmetric (conformal) theories that preserve some amount of superconformal symmetry, as this enables tractable computations and imposes constraints on observables.

Holographically, backgrounds dual to codimension-2 conformal monodromy defects preserving some amount of supersymmetry have been analysed in \cite{Arav:2024exg,Arav:2024wyg,Gutperle:2022pgw,Gutperle:2023yrd,Conti:2025wwf,Conti:2025wyj,Bomans:2024vii,Gutperle:2018fea,Gutperle:2019dqf,Chen:2020mtv,Capuozzo:2023fll} (in all these cases the ambient theory also preserves conformal symmetry). Such solutions are obtained at the level of lower dimensional gauged supergravities, and they share two important features: First, the metric asymptotes to $\AdS_{d}$ written as a foliation of $\AdS_{d-2}\times \Sp{1}$ over an interval. Second, the gauge fields have a non-trivial holonomy around the $\Sp{1}$. These are dual to background gauge fields for the abelian subgroup of the global symmetries, and the non-trivial holonomy of the supergravity gauge field leads to a non-trivial monodromy around the defect on the field theory.

Here, we focus on codimension-2 supersymmetric monodromy defects in ($p$+1)-dimensional maximally supersymmetric SU($N$) Yang-Mills (YM) for\footnote{The $p=3$ case has already been studied in \cite{Arav:2024exg,Conti:2025wwf}, but we review it here for completeness.} $p=2,3,4$. We note that for $p\neq3$ neither the ambient theory nor the defect preserve conformal symmetry. These solutions are obtained by considering the D$p$-branes wrapping spindles of  \cite{Ferrero:2024vmz}, and changing the values of the integration constants, such that the coordinate domain of one of the coordinates of the spindle is now semi-infinite. This procedure was also applied in \cite{Capuozzo:2023fll,Gutperle:2022pgw,Gutperle:2023yrd,Bomans:2024vii,Conti:2025wwf,Conti:2025wyj}.

A key element of our analysis is that the brane frame metric of the decoupling limit of a stack of D$p$-brane (for $p\neq5$) is AdS \cite{Skenderis:1998dq,Kanitscheider:2008kd}. Although there is no conformal symmetry, as the dilaton depends on the AdS radial coordinate, the fact that the metric in this frame is AdS, allows us to extend the same techniques used in the conformal cases mentioned above to the ones studied here. The case $p=5$ is analysed separately.  

\subsection{Brief Comments on Monodromy Defects}\label{sec:FieldTheoryDefects}

As mentioned above, we are interested in describing codimension-2 monodromy defects in $(p+1)$-dimensional maximally supersymmetric SU($N$) YM on flat space time. These theories have a SO($9-p$) global R-symmetry, which has U(1)$^{\mathfrak{r}}$ abelian subgroup, with $\mathfrak{r}$ the rank of SO($9-p$), $\mathfrak{r}= \floor{\frac{9-p}{2}}$. In the cases of our interest, which preserve at least two supercharges,  this abelian subgroup is split into a U(1)$_{\RR}$ symmetry and a U(1)$_{\FF}^{\mathfrak{r}-1}$ flavour symmetry. To introduce codimension-2 monodromy defects for these symmetries, we write the $(p+1)$-dimensional metric with two space directions written in polar coordinates
    \begin{equation}\label{eq:FlatMetricIntro}
        \dd s^{2}_{p+1} = \dd x^{2}_{1,p-2} + \dd \hat{\rho}^{2} + \hat{\rho}^{2} n^{2}\dd z^{2},
    \end{equation}
where $z\sim z+2\pi$, and we have allowed for conical singularities via the parameter $n$. We show in Section \ref{sec:Defects} that for the theories under consideration, it is only possible to insert a \textit{supersymmetric} monodromy defect for the U(1)$_{\RR}$ when $n\neq1$, which has also been observed in superconformal theories \cite{Arav:2024exg,Arav:2024wyg,Conti:2025wwf,Conti:2025wyj}. A defect located at $\hat{\rho}=0$ is inserted by coupling the theory to a background gauge field for the  U(1)$_{\RR} \times $U(1)$_{\FF}^{\mathfrak{r}-1}$ symmetries, as \cite{Bianchi:2021snj}
    \begin{equation}
        \mathcal{A}_{\RR} = \mu_{\RR}\, \dd z, \quad
        \mathcal{A}_{\FF_{\hat{I}}} = \mu_{\hat{I}}\,  \dd z,
    \end{equation}
with $\hat{I}=1,...,\mathfrak{r}-1$. This is achieved by modifying the gauge covariant derivative $D$ as
    \begin{equation}
        D_{\mu}\Psi \rightarrow \calD_{\mu}\Psi = D_{\mu}\Psi - i \calR[\Psi] (\calA_{\RR})_{\mu}\Psi - i \calF_{\hat{I}}[\Psi] (\calA_{\FF_{\hat{I}}})_{\mu}\Psi,
    \end{equation}
where $\calR[\Psi]$ and $\calF_{\hat{I}}[\Psi]$ are the charges of $\Psi$ under the U(1)$_{\RR}$ and U(1)$_{\FF_{\hat{I}}}$ symmetries respectively. 

Since these gauge fields are singular at $\hat{\rho}=0$, the SO($1,p$) Lorentz symmetry is broken\footnote{If both the ambient theory and the defect also preserve conformal symmetry, as is the case for $p=3$, then it is broken as SO($2,p+1$)$\rightarrow$ SO($2,p-1$) $\times$ SO(2).} to SO($1,p-2$) $\times$ SO(2), as expected for a codimension-2 defect. The effect of this background gauge field is to make fields charged under these symmetries pick-up a non-trivial monodromy around the defect. Consider for example the gaugino ($\lambda$), which is only charged under the U(1)$_{R}$. Under $z\rightarrow z+2\pi$, it picks-up an extra factor of $e^{-i \mu_{\RR} \calR[\lambda] }$.

Finally, in order to make an easier connection with the supergravity solutions, we recall that the flat metric \eqref{eq:FlatMetricIntro} is conformally AdS$_{p}\times \Sp{1}$ 
    \begin{equation}\label{eq:IntroWeylRescaling}
        \dd s^{2}_{p+1} = \hat{\rho}^{2}\left( \frac{\dd x^{2}_{1,p-2}+\dd \hat{\rho}^{2}}{\hat{\rho}^{2}} + n^{2} \dd z^{2}\right).
    \end{equation}  
To perform a Weyl transformation of the metric, we recall that maximally supersymmetric Yang-Mills preserves a \textit{generalized conformal structure} \cite{Kanitscheider:2008kd}: the curved space lagrangian\footnote{By this, we mean the SYM lagriangian in a generic curved spacetime, which has the spacetime covariant derivative for the fermions and the $R \Phi^{2}$ coupling} is invariant under a Weyl transformation, under which the background metric $g_{\mu\nu}$, the ($9-p$) adjoint scalars $X^{I}$, the adjoint fermions $\psi^{a}$ and the gauge coupling $g_{\YM}$ transform as  
    \begin{equation}
        g_{\mu\nu} \rightarrow \Omega^{2} g_{\mu\nu}, \quad
        X^{I} \rightarrow \Omega^{-\frac{p-1}{2}} X^{I},\quad
        \psi^{a} = \Omega^{-\frac{p}{2}} \psi^{a},\quad
        g_{\YM} = \Omega^{\frac{p-3}{2}} g_{\YM}.
    \end{equation}
Choosing $\Omega = \hat{\rho}^{-1}$ leads to maximally supersymmetric Yang-Mills formulated on AdS$_{p}\times\Sp{1}$. We note that, while in the $p=3$ case the gauge coupling does not change, in the non-conformal cases $p\neq3$, the Weyl transformation leads to a position dependent gauge coupling. For example, focusing on the Yang-Mills kinetic term, which after the Weyl rescaling reads
    \begin{equation}\label{eq:IntroDBI}
        S_{\YM} = \int \dd ^{p+1}x\, \sqrt{-\tilde{g}}\, \left( -\frac{\hat{\rho}^{p-3}}{2g^{2}_{\YM}} \tilde{g}^{\mu\nu}\tilde{g}^{\lambda\rho}\, \text{Tr}(F_{\mu\lambda}F_{\nu\rho}) \right),
    \end{equation}
the would be gauge coupling depends on the AdS$_{p}$ radial coordinate $\hat{\rho}$. We show below that this effective gauge coupling is matched by holography. Although we do not perform supersymmetry analysis on the field theory side, we explain via holography in Section \ref{sec:Dp-branes}, why this Weyl rescaling preserves maximal supersymmetry of the vacuum theory and does not change the SO($9-p$) R-symmetry.

In the conformal case, $\AdS_{p}\times\Sp{1}$ precisely realises the spacetime symmetries of a codimension-2 \textit{conformal} defect, while in the non-conformal case, the factor of $\hat{\rho}^{p-3}$ breaks the $\AdS_{p}$ isometries to SO($1,p-2$), matching the discussion above. 

Finally, we note that after the Weyl rescaling there is no defect in field theory. Instead, we have maximally supersymmetric SU($N$) Yang-Mills on AdS$_{p}\times\Sp{1}$ (the position dependent gauge coupling is necessary to preserve maximal supersymmetry). In this Weyl frame, the constant gauge field around the $\Sp{1}$ realises a twisted circle compactification of the theory, as the ones described in \cite{Kumar:2024pcz} for theories in $\mathbb{R}^{1,p-1}\times\Sp{1}$.

\subsection{Organization of the paper}

This paper is organized as follows: in Section \ref{sec:Dp-branes} we start by reviewing general aspects of near horizon geometries to D$p$-branes and computing the central charge / free energy of the dual theories. We quickly review the brane frame of \cite{Skenderis:1998dq,Kanitscheider:2008kd} and find coordinates adapted to the study of defect solutions. In Section \ref{sec:Defects}, we define the defect boundary conditions as in \cite{Arav:2024exg}. For $p=2,3,4$, we change the coordinate range of the spindle solutions in \cite{Boisvert:2024jrl} to interpret them as dual to codimension-2 defects. Section \ref{sec:p5} is dedicated to the study of defects on D5-branes, which do not enjoy an AdS metric in brane frame. Then, in Section \ref{sec:DefectsEE} we provide a prescription to compute the defect entanglement entropy. This amounts to interpreting the supergravity solutions as dual to a ($p-1$)-dimensional theory and applying the formulas of \cite{Macpherson:2014eza,Bea:2015fja}. To renormalize a divergence, we define a UV cut-off, for which the conformal AdS behaviour of the D$p$-brane ($p<5$) plays a central role. For $p=2,3,4$ we find that the defect entanglement entropy is proportional to the free energy of the ambient theory. This generalizes the results of \cite{Conti:2025wwf} for conformal monodromy defects to non-conformal ones. We conclude in Section \ref{sec:Conclusion} with a summary of our results and comments on lines of future research. An appendix with the supergravities used throughout this paper is provided.

\section{D\texorpdfstring{$p$}{p}-branes and the Dual Frame}\label{sec:Dp-branes}

In this section we review the Type II backgrounds holographically dual to ($p$+1)-dimensional maximally supersymmetric SU($N$) Yang-Mills theories. These correspond to the decoupling limit of a stack of $N$ D$p$-branes. It is convenient to rewrite these solutions using the change of coordinates of \cite{Skenderis:1998dq,Kanitscheider:2008kd} that makes the metric conformally $\AdS_{p+2}\times \Sp{8-p}$. It is this fact that allows us to generalize the study of codimension-2 monodromy defects in conformal theories to the maximally supersymmetric Yang-Mills theories in dimensions in which the theory is not conformally invariant. We also discuss a change of coordinates that matches the asymptotic behaviour of the monodromy defect solutions studied in the next section.

We start by considering the background configuration describing the backreaction of a single stack of $N$ D$p$-branes, which in string frame is given by\footnote{For the D3-brane, this only gives the magnetic part of the self-dual 5-form. To solve the equations of motion, self-duality has to be imposed.}
    \begin{subequations}
    \begin{align}
        \dd s^{2}_{\text{st}} &=  H_{p}(r)^{-\frac{1}{2}}\dd x^{2}_{1,p} + H_{p}(r)^{\frac{1}{2}}\left( 
                   \dd r^{2} + r^{2} \dd s^{2}\left( \Sp{8-p}\right) \right), \label{eq:DpMetric} \\[2mm]
        F_{8-p} &= (7-p) \ell^{7-p} \vol\left(\Sp{8-p}\right),\label{eq:DpFlux} \\[2mm]
        e^{\Phi} &=  H_{p}(r)^{\frac{3-p}{4}}\label{eq:DpDilaton},
    \end{align}
    \end{subequations}
where $\vol(\Sp{8-p})$ is the volume form of the $(8-p)$-sphere, and in the decoupling limit
    \begin{equation}\label{eq:HandL}
        H_{p}(r) = \frac{\ell^{7-p}}{r^{7-p}}, \qquad 
        \ell^{7-p} = (4\pi)^{\frac{5-p}{2}} \Gamma\left( \frac{7-p}{2} \right) N\, (\alpha')^{\frac{7-p}{2}}g_{s}.
    \end{equation}
There are $N$ units of D$p$-brane flux through the $\Sp{8-p}$
    \begin{equation}
        \frac{1}{(2\pi\sqrt{\alpha'})^{7-p}g_{s}}\int_{\Sp{8-p}}F_{8-p} = N.
    \end{equation}

It was argued in \cite{Skenderis:1998dq,Kanitscheider:2008kd} that the metric \eqref{eq:DpMetric} is conformally $\AdS_{p+2}\times \Sp{8-p}$. This can be seen explicitly using the change of coordinates
    \begin{equation}\label{eq:CoordChangeBraneFrame}
        \left(\frac{r}{\ell}\right)^{5-p} = \calR^{2}\, (u\, \ell)^{2}, \qquad
        \calR^{2} = \left(\frac{2}{5-p}\right)^{2}. 
    \end{equation}
This change of coordinates exists for D$p$-branes for $p\leq 4$, we study the D5-brane case separately. In these coordinates, the boundary is located at $u\rightarrow +\infty$ and the metric and dilaton are
    \begin{subequations}
    \begin{align}
        \dd s^{2}_{\text{st}} &= \left( \calR^{2} (u\, \ell)^{2} \right)^{-\frac{p-3}{2(p-5)}} \ell^{2} \left(\calR^{2} \left(  u^{2} \dd x^{2}_{1,p} + \frac{\dd u^{2}}{u^{2}} \right)  +   \dd s^{2}\left(\Sp{8-p}\right) \right), \label{eq:MetricBraneFrame} \\
        e^{\Phi} &= \left( \calR^{2} \, (u\,\ell)^{2} \right)^{\frac{(p-3)(p-7)}{4(p-5)}}\label{eq:DilatonBraneFrame}.
    \end{align}
    \end{subequations} 

In the Brane frame\footnote{We change conventions with respect to \cite{Skenderis:1998dq,Kanitscheider:2008kd}.}, also called dual frame for non-conformal D$p$-branes
    \begin{equation}\label{eq:StringToDualFrame}
        \dd s^{2}_{\dual} = \left(\ell^{7-p}e^{\Phi}\right)^{-\frac{2}{7-p}}\dd s^{2}_{\text{st}}, \\
    \end{equation}
it was argued \cite{Skenderis:1998dq,Boonstra:1998mp} that the coordinate $u$ is identified with the energy scale of the boundary theory. In fact, recalling that the effective (dimensionless) 't Hooft coupling in $(p+1)$ dimensions is given in terms of the energy scale $E$ as
    \begin{equation}\label{eq:Defthooft}
        \lambda_{\text{eff}} = g^{2}_{\YM}\,N E^{p-3}, \qquad
        g^{2}_{\YM} = g_{s} (2\pi)^{p-2} (\alpha')^{\frac{p-3}{2}} \\
    \end{equation}
with $g^{2}_{\YM}$ the Yang-Mills coupling constant, using the identification $u\sim E$, we can write a relation between the dilaton and the effective 't Hooft coupling\footnote{Here, the factor of $g_{s}$ on the right hand side is added manually when compared to \cite{Kanitscheider:2008kd}, since $e^{\Phi_{\text{there}}} = g_{s} e^{\Phi_{\text{here}}}$.}
    \begin{equation}\label{eq:DilatontHooft}
        N\, g_{s} e^{\Phi} = c_{d} \lambda_{\text{eff}}^{\frac{7-p}{2(5-p)}}, \\
    \end{equation}
where $c_{d}$ can be found in equation (2.21) of \cite{Kanitscheider:2008kd}. In this frame it is also possible to read the boundary metric to be the flat $(p+1)$-dimensional Minkowski one. The  gauge coupling  can be determined from the DBI action, which we write using the dual frame metric
    \begin{equation}\label{eq:DBIBraneFrame}
        S_{\text{DBI}} = \frac{\ell^{p-3}}{(2\pi)^{p-2}(\alpha')^{\frac{p-3}{2}}g_{s}} \int \dd ^{p+1}x\, \sqrt{-\det(g_{\text{d,Ind}})}\, \left(e^{\Phi}\right)^{\frac{2(p-5)}{7-p}}\left(  -\frac{1}{4}F_{\mu\lambda}F_{\nu\rho} (g_{\text{d,Ind}})^{\mu\nu}(g_{\text{d,Ind}})^{\lambda\rho}\right),
    \end{equation}
where $g_{\text{d,Ind}}$ is the induced dual frame metric on the $(p+1)$-dimensional Minkowski space. From \eqref{eq:StringToDualFrame} and \eqref{eq:MetricBraneFrame}-\eqref{eq:DilatonBraneFrame} we read $g_{\text{d,Ind}} = \calR^{2}u^{2}\, \eta$, with $\eta$ the flat Minkowski metric. In \eqref{eq:DBIBraneFrame}, the gauge coupling is read by identifying the coefficient of $F^{2}$ with $1/(4g^{2}_{\YM})$. In this case, using \eqref{eq:DilatontHooft} one recovers \eqref{eq:Defthooft}, as was pointed out in \cite{Kanitscheider:2008kd}.

The free energy\footnote{We note that for even dimensional conformal theories the free energy matches the Weyl anomaly, which in holographic CFTs it is also equal to the central charge.} of these theories can be computed using \cite{Macpherson:2014eza,Bea:2015fja} (this is computed using the string frame metric, see Section \ref{sec:RenormScheme} for details) 
    \begin{equation}\label{eq:CCp+1}
        c_{\hol}^{(p)} = \frac{2}{G_{N}} \frac{p^{p}\pi^{\frac{9-p}{2}}}{\Gamma\left( \frac{9-p}{2} \right)\left(9-p\right)^{p}} \left( 
            \calR^{2} (u\,\ell)^{2} \right)^{-\frac{(p-3)^{2}}{2(p-5)}} \ell^{8}, \\
    \end{equation}
where we have added the superscript $(p)$ as a label. We can write this expression in terms of the effective 't Hooft coupling and $N$, using \eqref{eq:DilatonBraneFrame} and \eqref{eq:DilatontHooft}. Omitting numerical factors, we find the free energy to be
    \begin{equation}
        c_{\hol}^{(p)} \sim N^{2} \lambda_{\text{eff}}^{\frac{p-3}{5-p}}, \\
    \end{equation}
which matches the supersymmetric localization computation \cite{Minahan:2015any} and its holographic version using spherical branes \cite{Bobev:2018ugk,Bobev:2019bvq}. We note that for D$p$-brane solutions, the statement above is true independently of the choice of parametrization for the holographic coordinate.

The advantage of using the parametrization in \eqref{eq:MetricBraneFrame}-\eqref{eq:DilatonBraneFrame} where there is an explicit AdS factor, even if there is no manifest AdS isometry in the background, is that we can apply a similar analysis to those in \cite{Arav:2024exg,Arav:2024wyg,Conti:2025wwf} for codimension-2 monodromy defects in AdS$_{d+2}$ geometries. The vacuum of these solutions corresponds to AdS$_{d+2}$ written as a foliation of an interval over $\AdS_{d}\times \Sp{1}$, and the defect solution is obtained as deformations of this vacuum. It is convenient for us to parametrize the D$p$-brane solutions above in this way and to discuss some general features of this coordinate choice, since the defect solutions studied in the following sections asymptote to the vacuum written in this form.

We can obtain such a parametrization of the pure D$p$-brane geometries, which correspond to the vacuum of the solutions studied in Section \ref{sec:Defects} below, that is, we move to coordinates adapted to the study of codimension-2 defect solutions by first writing the Minkowski metric as
    \begin{equation}\label{eq:PolarMink}
        \dd x^{2}_{1,p} = \dd x^{2}_{1,p-2} +  \dd \tilde{y}^{2} + \tilde{y}^{2}\dd z^{2}, \\
    \end{equation}
then, we perform a change of coordinates that mixes the holographic coordinate $u$ and the coordinate $\tilde{y}$ belonging to the field theory directions, $(u,\tilde{y}) \rightarrow (\rho,y)$
    \begin{equation} \label{eq:ChangeToDefect}
        u = \rho y, \qquad
        \tilde{y}^{2} = \frac{1}{\rho^{2} y^{2}}(y^{2}-1), \\
    \end{equation}
where $\rho>0$ and $y>1$. In these coordinates, for $p=2,3,4$ the AdS boundary (UV region) $u\rightarrow +\infty$ is mapped to $y\rightarrow+\infty$. The D$p$-backgrounds in dual frame now reads
    \begin{subequations}
    \begin{align}
        \dd s^{2}_{\dual} &= \calR^{2}y^{2} \left( \rho^{2}\dd x^{2}_{1,p-2}  + \frac{\dd \rho^{2}}{\rho^{2}} + \frac{1}{y^{2}(y^{2}-1)}\dd y^{2} + \frac{y^{2}-1}{y^{2}}\dd z^{2}\right) + \dd s^{2}(\Sp{8-p}) , \label{eq:NewDpMetric} \\[2mm]
        F_{8-p} &= (7-p) \ell^{7-p} \vol\left(\Sp{8-p}\right),\label{eq:NewDpFlux} \\[2mm]
        e^{\Phi} &= \left(\calR^{2}\, (y\, \rho \, \ell)^{2}  \right)^{\frac{(p-7)(p-3)}{4(p-5)}}\label{eq:NewDpDilaton},
    \end{align}
    \end{subequations}
Near $y=1$ the geometry shrinks smoothly\footnote{We note that under $y  = \sec(\theta)$, 
    \begin{equation}
        \frac{1}{y^{2}(y^{2}-1)}\dd y^{2} + \frac{y^{2}-1}{y^{2}}\dd z^{2} 
        = \dd \theta^{2} + \sin^{2}(\theta)\dd z^{2}.
    \end{equation}
Although the metric looks like a $\Sp{2}$, there is no SO(3) isometry since the warp factors and the dilaton depend on $\theta$.}
    \begin{equation}
        \frac{1}{y^{2}(y^{2}-1)}\dd y^{2} + \frac{y^{2}-1}{y^{2}}\dd z^{2}   \underset{y\rightarrow 1}{\longrightarrow} \dd y^{2} + y^{2} \dd z^{2}. \\
    \end{equation}
On the other hand, we can read the boundary metric at the UV region $y\rightarrow +\infty$, from the asymptotic expansion (here we use $\rho = 1/\hat{\rho}$)
    \begin{equation}
        \dd s^{2} \rightarrow \frac{\dd y^{2}}{y^{2}} 
        +  y^{2} \left( \frac{1}{\hat{\rho}^{2}} \left( \dd x^{2}_{1,p-2}  + \dd \hat{\rho}^{2} \right) + \dd z^{2}\right) , \\
    \end{equation}
so that the boundary metric is $\AdS_{p}\times\Sp{1}$, which is conformally $(p+1)$-dimensional Minkowski
    \begin{equation}\label{eq:yatinfinity}
        \frac{1}{\hat{\rho}^{2}} \left( \dd x^{2}_{1,p-2}  + \dd \hat{\rho}^{2}  \right) + \dd z^{2} = \frac{1}{\hat{\rho}^{2}} \left( \dd x^{2}_{1,p-2}  + \dd \hat{\rho}^{2}  + \hat{\rho}^{2}\dd z^{2} \right). \\
    \end{equation}
This exact behaviour is found in the conformal defect solutions of \cite{Arav:2024exg,Arav:2024wyg,Conti:2025wwf}. In that context, the holographic interpretation is that one takes a Weyl transformation of the flat Minkowski metric, so that the theory is now formulated in $\AdS_{p}\times \Sp{1}$. In this case, the interpretation is similar, but since for $p=2,4$ the theories are non-conformal, the Weyl rescaling also induces a non-trivial profile for the gauge coupling, which can be read using the DBI action \eqref{eq:DBIBraneFrame} to be\footnote{This corresponds to the boundary value $y\rightarrow +\infty$. The leading coefficient in the asymptotic expansion corresponds to the gauge coupling \cite{Kanitscheider:2008kd}.} $\hat{\rho}^{p-3}/g^{2}_{\YM}$, which coincides with the field theory discussion around \eqref{eq:IntroDBI}. We also note that the flat space axis located at $\hat{\rho}=0$ is mapped to infinity after the Weyl rescaling.

We note that since the coordinate change \eqref{eq:ChangeToDefect} is regular, it does not change the amount of supersymmetry preserved by the background. This, combined with the fact that the $\Sp{8-p}$ is not deformed by the change of coordinates, implies that the background$-$and hence the dual theory$-$preserves the same amount supersymmetry. Although this confirms the claim of Section \ref{sec:FieldTheoryDefects}  that it is possible to preserve maximal supersymmetry on AdS$_{p}\times \Sp{1}$ by allowing the gauge coupling to be position-dependent, a purely field theory analysis along the lines of \cite{Festuccia:2011ws,Dumitrescu:2012ha,Maxfield:2016lok} would be valuable to further confirm this claim. We hope to report on this in the near future.

To obtain the background dual to a codimension-2 monodromy defect in maximally supersymmetric Yang-Mills in $(p+1)$ dimensions, we can extend the procedure of \cite{Capuozzo:2023fll,Gutperle:2022pgw,Gutperle:2023yrd,Bomans:2024vii,Conti:2025wwf,Conti:2025wyj} to non-conformal branes. This is, we start by considering the solutions from \cite{Boisvert:2024jrl} of D$p$-branes ($p=2,4,5$) wrapped on spindles\footnote{The $p=3$ case was studied in \cite{Capuozzo:2023fll,Gutperle:2022pgw,Gutperle:2023yrd,Bomans:2024vii,Conti:2025wwf,Conti:2025wyj}, but we briefly review it for consistency.}. Then, by considering a different range of the spindle coordinates, we can interpret the solution as a codimension-2 monodromy defect. We discuss this in detail in Section \ref{sec:Defects} for $p=2,3,4$ and Section \ref{sec:p5} for $p=5$.

\section{Codimension-2 Monodromy Defects for \texorpdfstring{$p=2,3,4$}{p=2,3,4}}\label{sec:Defects}

We now proceed to the holographic study of codimension-2 monodromy defects in ($p$+1)-dimensional maximally supersymmetric SU($N$) YM. To obtain these solutions, we follow \cite{Capuozzo:2023fll,Gutperle:2022pgw,Gutperle:2023yrd,Bomans:2024vii,Conti:2025wwf,Conti:2025wyj}, that is, we start by considering the D$p$-branes wrapped on spindle solutions of \cite{Boisvert:2024jrl} and consider a different range of the spindle coordinates, which allows us to interpret the solutions as defects.

\subsection{General Aspects of the Solutions and Boundary Conditions}\label{sec:GeneralDefect}

We start by considering the solutions of D$p$-branes wrapped on a spindle of \cite{Boisvert:2024jrl}. These are solutions of a U(1)$^{\mathfrak{r}}$ truncation of the $(p+2)$-dimensional SO($9-p$) gauged supergravity, here $\mathfrak{r}$ is the rank of SO($9-p$), obtained by reducing the 10D theory\footnote{For $p=2$, that is, the reduction on $\Sp{6}$, the resulting 4D gauged supergravity has ISO(7) as gauge group. Here we consider only the compact SO(7) subgroup.} on $\Sp{8-p}$. The bosonic field content of these lower dimensional U(1)$^{\mathfrak{r}}$ theories are: the metric, $\mathfrak{r}$ U(1) gauge fields and a proper number of scalars whose number depends on $p$. The metric of the solutions under consideration takes the form\footnote{We perform a change of coordinates of the form $y\rightarrow y^{c}$ of the solutions of \cite{Boisvert:2024jrl} and a redefinition of $H$ and $P$ such that all the backgrounds behave in the way presented here.}
    \begin{equation}\label{eq:GeneralMetric}
        \dd s^{2}_{p+2} = F(\rho,y)\left( \rho^{2}\dd x^{2}_{1,p-2} + \frac{\dd \rho^{2}}{\rho^{2}} + \dd s^{2}(\Sigma) \right), \\
    \end{equation}
where
    \begin{equation}\label{eq:Spindle}
        \dd s^{2}(\Sigma) = \frac{1}{P} \dd y^2 + \frac{P}{H} n^{2} \dd z^{2}, \\
    \end{equation}
with $z$ a U(1) isometry of period 2$\pi$, while $n$ parametrizes the angular deficit ($n<1$) or excess ($n>1$) and $P$ and $H$ satisfy $P>0$ and $H>0$. The gauge fields and scalars take the form
    \begin{equation}\label{eq:GeneralGaugeScalar}
        A^{I} = a^{I}(y)\, \dd z, \quad \Phi^{(p)} = \Phi^{(p)}(\rho,y), \\
    \end{equation}
where $I = 1,...,\mathfrak{r}$ and $\Phi^{(p)}$ denotes the set of scalars of the ($p+2$)-dimensional theory. The explicit form of $F$, $H$, $P$, $a^{I}$ and the scalars depends explicitly on $p$ (we explicitly show the solutions below). We note that the supersymmetric solutions of the U(1)$^{\mathfrak{r}}$ theories under consideration are characterized by  $2\mathfrak{r}$ integration constants $(q_{I},\alpha_{I})$.

In the case of \cite{Boisvert:2024jrl}, the values of the integration constants were chosen such that the range of $y$ was taken between zeros of the function $P$, denoted $y_{1}$ and $y_{2}$, that is $P(y)>0$ for $y_{1}<y<y_{2}$. Then, regularity conditions are imposed such that \eqref{eq:Spindle} corresponds to a spindle. 

In order to interpret the solutions of \cite{Boisvert:2024jrl} as defects, we follow the procedure of \cite{Capuozzo:2023fll,Gutperle:2022pgw,Gutperle:2023yrd,Bomans:2024vii,Conti:2025wwf,Conti:2025wyj}: we extend the range of $y$ to be semi-infinite, that is, $y\in [y_{\core}, +\infty)$, where $y_{\core}$ is the greatest zero of the function $P$. In order to achieve this, we need to change the possible values of the integration constants. Furthermore, we need to impose defect boundary conditions as in \cite{Arav:2024exg,Arav:2024wyg,Conti:2025wwf,Conti:2025wyj}. 
    \begin{itemize}
        \item We allow the metric and scalars to be singular at $\rho=0$ and $\rho\rightarrow +\infty$, since these singularities are already present in the vacuum solutions \eqref{eq:NewDpMetric}-\eqref{eq:NewDpDilaton}.
        \item At $y=y_{\core}$, we require \eqref{eq:Spindle} to shrink smoothly. For this, besides $P(y_{\core})=0$ we require $P'(y_{\core})\neq0$ and $H(y_{\core})\neq0$. Then, regularity of the 2D space $\Sigma$ at $y=y_{\core}$ leads to 
        \begin{equation}\label{eq:reg2dmetric}
            \dd s^{2}(\Sigma)\bigg|_{y=y_{\core}}  \rightarrow  \dd y^{2} + k^{2} y^{2} \dd z^{2}, \quad k^{2} = \frac{(P'(y_{\core}))^{2}}{4H(y_{\core})}n^{2}. \\
        \end{equation}
        Since $z\sim z +2\pi$, one must impose $k^{2}=1$. Also, we impose that the gauge fields $A^{I}$ vanish at the end of the space
        \begin{equation}\label{eq:reggauge}
            A^{I}(y_\core) = 0,  \\
        \end{equation}
    and scalars to be regular.
    \item When $y\rightarrow +\infty$, the metric (in dual frame) asymptotes to the vacuum \eqref{eq:yatinfinity}, with $\dd z^{2}\rightarrow n^{2}\dd z^{2}$. If $n\neq 1$ there is either a conical deficit or excess in the boundary theory. For the gauge fields we have
        \begin{equation} \label{eq:defmuI}
        \frac{1}{2\pi}\int_{\Sp{1}} A^{I} \bigg|_{y\rightarrow +\infty} = \mu^{I}. \\
    \end{equation}
    Here, the parameters $\mu^{I}$ are interpreted as background gauge fields for the U(1)$^{\mathfrak{r}}$ subgroup of the R-symmetry of the dual maximal theory. Importantly, $\mu^{I}\dd z$ cannot be gauged away. As in the conformal defects of \cite{Arav:2024exg,Arav:2024wyg}, only for $\mu^{I} \neq0$ these solutions describe the insertion of a codimension-2 monodromy defect in the dual ($p+1$)-dimensional theory, located at $\hat{\rho}=0$ in flat space. However, in the cases studied here (for $p\neq3$), neither the ($p+1$) theory nor the defect are conformally invariant. 

    After the Weyl rescaling that sets the field theory metric to be $\AdS_{p}\times \Sp{1}$, discussed below \eqref{eq:yatinfinity} for the vacuum, the background gauge fields $\mu^{I}dz$ have a non-trivial holonomy around the $\Sp{1}$. Also, similar to the conformal defects of \cite{Arav:2024exg,Arav:2024wyg,Conti:2025wwf}, in the solutions studied below the conical singularity parameter $n$ is only related to the monodromy of the gauge field dual to the U(1) R-symmetry. 
    \end{itemize}

We now proceed to the study of the $(p+2)$-dimensional solutions of \cite{Boisvert:2024jrl}, imposing the defect boundary conditions. The procedure is systematic and it is as follows: first we impose \eqref{eq:reggauge} and \eqref{eq:defmuI}, which allows to find an expression for the integration constants $q_{I}$ in terms of $y_{\core}$ and $\mu^{I}$, that is
    \begin{equation}\label{eq:SolvingqI}
        A^I(y_{\core}) = 0 \quad \text{and} \quad 
       \frac{1}{2\pi}\int_{\Sp{1}} A^{I} \bigg|_{y\rightarrow +\infty} = \mu^{I} \quad 
       \Rightarrow \quad 
       q_{I} = q_{I}(y_{\core},\mu^{I}), \\
    \end{equation}
then, using this expression for $q_{I}$ in the condition $P(y_{\core})=0$, allows to obtain an expression of $y_{\core}$ in terms of $\mu^{I}$
    \begin{equation}\label{eq:Solvingycore}
        P(y_{\core}) = 0 \quad \text{and} \quad q_{I} = q_{I}(y_{\core},\mu^{I}) \quad
        \Rightarrow \quad
        y_{\core} = y_{\core}(\mu^{I}). \\
    \end{equation}
Finally, imposing $k^{2}=1$ in \eqref{eq:reg2dmetric} leads to a constraint for the chemical potentials $\mu^{I}$.

\subsection{4D Gauged Supergravity} \label{sec:4D}

The first configuration we study corresponds to the D2-brane wrapping a spindle of \cite{Boisvert:2024jrl}\footnote{We use conventions such that the gauge coupling $g_{\text{here}}= \frac{1}{2} g_{\text{there}}$.}, which is a solution of the 4D U(1)$^3$  truncation of the ISO(7) gauged supergravity \cite{Hull:1984yy}. Here, we slightly change conventions to match the form of the solutions in \eqref{eq:GeneralMetric}, such that the vacuum solution (in brane frame) takes the form \eqref{eq:NewDpMetric}-\eqref{eq:NewDpFlux}. The solution takes the form
\begin{equation} \label{eq:4Dsolution}
\begin{split}
    \dd s^2 & = \left(\frac{2 }{3 g} \right)^{7/3} y^{1/3} H^{1/2} \rho^{1/3} \left(- \rho^2 \dd t^2 + \frac{\dd \rho^2}{\rho^2} + \frac{\dd y^2}{P} + \frac{P}{H} n^2 \dd z^2 \right),\\[2mm]
    A^I & = \left( \alpha_I + \kappa \frac{2}{3 g} \sqrt{1-\frac{\nu}{q_I}} \frac{q_I}{h_I} n \right) \dd z, \\[2mm]
    e^{\phi_0} & = \left(\frac{2}{3 g} \right)^{2/3} y^{2/3} \rho^{2/3}, \qquad e^{\phi_I} = \left(\frac{2}{3 g }\right)^{1/3} \frac{\rho^{1/3} \sqrt{H}}{y^{1/3} h_I}, 
\end{split}
\end{equation}
where
\begin{equation}\label{eq:FunctionsD2brane}
    P = H - y^2 - \nu y^{2/3}, \qquad H= h_1 h_2 h_3, \qquad h_I = y^{4/3} + q_I.
\end{equation}
We have also included the BPS-parameter $\nu$ that completely breaks supersymmetry. Besides $\nu$, the configuration is characterized by six integration constants $(\alpha_I,q_I)$. Also, $n$ is a conical singularity parameter (recall $z\sim z+2\pi$), while $\kappa=\pm1$. The vacuum solution is obtained by setting $(\alpha_I,q_I)=0$ and $n=1$. We provide the uplift to Type IIA in Appendix \ref{ap:4DGaugedSugra}. 

In what follows we focus on the supersymmetric defect solutions, so we set $\nu=0$. As has been discussed in \cite{Ferrero:2024vmz}, for generic $q_I$ the solution preserves two supercharges, ie 1D ${\mathcal{N}}=2$. As we will describe later in more details, specific values of $q_I$ allow for enhanced supersymmetry. We also fix $\kappa=1$ for convenience, as it can always be set to this value by $z\rightarrow -z$.

As explained above, in order to interpret the solution as dual to a codimension-2 defect, we need to impose suitable boundary conditions. These have been explained in Section \ref{sec:GeneralDefect}. We start by imposing boundary conditions for the gauge field \eqref{eq:SolvingqI} and that $y_{\core}$ is a zero of $P$ \eqref{eq:Solvingycore}, from where
\begin{equation}\label{eq:ycore1st}
    q_I = -\frac{3 g \mu^I y_{\core}^{4/3}}{3 g \mu^I + 2 n}, \qquad y_{\core} = \frac{\sqrt{3 g \mu^1 + 2 n} \sqrt{3 g \mu^2 + 2 n} \sqrt{3 g \mu^3 + 2 n}}{\sqrt{8}n^{3/2}}. 
\end{equation}
Then, the next step is to impose a smoothly shrinking geometry at $y=y_{\core}$, as in \eqref{eq:reg2dmetric}. This, combined with \eqref{eq:ycore1st} leads to a constraint for the background gauge fields $\mu_{I}$
    \begin{equation}\label{eq:Rsymmreg4D}
    g \mu^1 + g \mu^2 + g \mu^3 = n-1 .
    \end{equation}
Using \eqref{eq:RFgaugefields4D}, we can rewrite this constraint and \eqref{eq:ycore1st} in terms of the background gauge field for the flavour symmetries $(\mu_{\FF_{1}},\mu_{\FF_{2}})$ and the R-symmetry $\mu_{\RR}$. First, we note that \eqref{eq:Rsymmreg4D} only fixes the R-symmetry background gauge field
    \begin{equation}\label{eq:4DbackgroundGaugeFieldR}
        g \mu_{\RR} = n-1.
    \end{equation}
From here, we see that a non-trivial background gauge field for the R-symmetry is only possible when there is a conical singularity in the dual field theory. This feature was already known for \textit{conformal} monodromy defects in \textit{conformal} theories \cite{Arav:2024exg,Arav:2024wyg}, and from \eqref{eq:4DbackgroundGaugeFieldR} we see that it also applies to \textit{non-conformal} defects in the maximally symmetric 3D theory, which is \textit{non-conformal}. 

Finally, we rewrite \eqref{eq:ycore1st} as explained above. We find that it only depends on the flavour background gauge fields and the conical singularity parameter
    \begin{equation}\label{eq:4dregy}
    y_{\core} = \frac{\sqrt{n+1 - g \mu_{\FF_1}-2 g \mu_{\FF_2} } \sqrt{n+1 - g \mu_{\FF_1} + g \mu_{\FF_2} } \sqrt{n+1 + 2 g \mu_{\FF_1} + g \mu_{\FF_2} }}{\sqrt{8} n^{3/2}}. 
    \end{equation}
We quickly present now some cases where the expressions get highly simplified.

\subsubsection*{U(1) Invariant Truncation}

This sub-sector is obtained by fixing $q_1 = q_2 = q_3$. We note that in this case both flavour gauge fields $A_{\FF_{1}} = A^{1}-A^{2}$ and $A_{\FF_{2}} = A^{2}-A^{3}$ are set to zero, so that the only non-trivial gauge field is the R-symmetry one. Therefore, as seen above, there is a defect only if we allow for $n \neq 1$. In this case the solution is characterized by
\begin{equation}
    y_{\core} = \frac{(1+n)^{3/2}}{2^{3/2} n^{3/2}}, \qquad q_1 = \frac{n^2-1}{4 n^2}, \qquad \alpha_1 = \mu^1 = \frac{1}{3g}(1-n).
\end{equation}
This sector preserves the same amount of supersymmetry as the general solution. We now comment on sub-sectors of enhanced supersymmetry.

\subsubsection*{Sectors with Enhanced Supersymmetry}

As argued in \cite{Ferrero:2024vmz}, there are points in the parameter space that show enhancement of supersymmetry. These points correspond to setting the gauge fields to zero. This can be understood by considering the uplift in Appendix \eqref{ap:4DGaugedSugra}: a trivial gauge field restores a subgroup of the isometries of the internal manifold. In Subsection 6.3.3 of \cite{Ferrero:2024vmz}, the authors argue that the $A^3 = 0$ case, that is achieved by setting $q_3 = 0$, doubles the supersymmetry to four supercharges, i.e. $D=1$ $\mathcal{N}=4$. This is very similar to what has already been observed for the 4D U($1$)$^4$ truncation of SO($8$) maximal gauged supergravity, see for example \cite{Duff:1999gh,Arav:2024wyg}.

We focus on the simplest case $A^{2}=A^{3}=0$. In this limit the internal manifold exhibits an SO(4) $\cong$ SU(2)$\times$ SU(2) isometry group, leading to a fourfold enhancement of supersymmetry, $D=1$ ${\mathcal{N}}=8$. In this case, the parameters defining the solution read
\begin{equation}
    y_{\core} = \frac{\sqrt{3-n}}{\sqrt{2}\sqrt{n}}, \qquad q_1 = \frac{3 (n-1)}{2^{2/3} (3-n)^{1/3} n^{2/3}}, \qquad \alpha_1 = \mu^1 = \frac{1}{g} (1-n).
\end{equation}

\subsection{5D Gauged Supergravity}

In this section we briefly review the 5D supergravity background dual to a codimension-2 defect on 4D $\mathcal{N}=4$ SU($N$) SYM. This solution was obtained in \cite{Kunduri:2007qy} and recently studied in \cite{Boido:2021szx,Hosseini:2021fge,Ferrero:2021etw} in the context of black hole near horizon geometries and spindle compactifications. The extension of this solution that allows to have a codimension-2 superconformal defect interpretation was investigated in \cite{Conti:2025wwf}, where this solution was analysed. We limit ourselves to presenting the solution and reviewing the main features, since it completes our discussion for codimension-2 defects in D$p$-brane theories. 

The background is a solution of the 5D U(1)$^{3}$ gauged supergravity. The bosonic fields are the metric, three gauge fields $A^{I}$ ($I=1,2,3$) and two scalars $\phi_{1}$ and $\phi_{2}$. See Appendix \ref{ap:5DGaugedSugra} for details. In the conventions of \eqref{eq:GeneralMetric}, the supergravity background reads 
    \begin{equation}
    \begin{split}\label{eq:5DU13solution}
        \dd s^2_5 & = \frac{y^{2/3} H^{1/3}}{g^2} \left[\dd s^2(\AdS_3) + \frac{1}{P} \dd  y^2+\frac{P}{H} n^2 \dd  z^2\right], \\[2mm]
        A^I & = \left( \alpha^I + \kappa \sqrt{1- \frac{\nu}{q_I}} \frac{n}{g}\frac{q_I}{h_I}\right) \dd z , \qquad 
        e^{\sqrt{6}\phi_{1}} = \frac{h_{1}h_{2}}{h_{3}^{2}}, \qquad
        e^{\sqrt{3}\phi_{2}} = \frac{h_{1}}{h_{2}}. \\
    \end{split}
    \end{equation}
Here, the $\AdS_{3}$ factor has unit radius and SO(2,2) is an exact isometry of the configuration. The functions $h_I$, $H$ and $P$ depend only on $y$ and are given by
    \begin{equation}\label{eq:5DU13functions}
        H = y^{-2} h_1 h_2 h_3 , \qquad 
        P = H - y^2 - \nu, \qquad
        h_I = y^2 + q_I. \\
    \end{equation}
As before, $\nu$ is a non-supersymmetric parameter and $\kappa = \pm 1$. Furthermore, using the uplift in Appendix \ref{ap:5DGaugedSugra} and imposing flux quantization we find $g=1/\ell$, with $\ell$ given by \eqref{eq:HandL} with $p=3$.

We focus on the supersymmetric case , so we set $\nu=0$. For general $q_I$ this solution preserves 2D $\mathcal{N}=(0,2)$ supersymmetry, as proved in \cite{Ferrero:2021etw}. There exist specific tuning of $q_I$ that allow to have an enhancement of supersymmetry, we will come back to this point in more detail later.

We now study defect boundary conditions. Using \eqref{eq:SolvingqI} and \eqref{eq:Solvingycore} we find
    \begin{equation}
        q_{I} = - \frac{g\, y^{2}_{\core} \mu^{I}}{n+g\, \mu^{I}}, \quad
        y_{\core} = \frac{1}{n^{\frac{3}{2}}} (n+g\mu^{1})^{\frac{1}{2}}(n+g\mu^{2})^{\frac{1}{2}}(n+g\mu^{3})^{\frac{1}{2}},
    \end{equation}
where we set $\kappa=1$ for convenience. Imposing now \eqref{eq:reg2dmetric} on our expression of $y_{\core}$ leads us to
\begin{equation}
    g \mu_1 + g \mu_2 + g \mu_3 = n-1, \quad \Rightarrow \quad 
    g \mu_{\RR} = n-1 ,
\end{equation}
where we have used the definition of the R-symmetry gauge field in \eqref{eq:RFgaugefields5D}. In terms of the flavor gauge field, we can write
\begin{equation}
    y_{\core} = \frac{\sqrt{2 (2 n+1)-3 g \mu_{\FF'}} \sqrt{4 (2 n+1)+3 g (\mu_{\FF'}-2 \mu_{\FF})} \sqrt{ 4 (2 n+1) +3 g (2 \mu_\FF + \mu_{\FF'}) }}{12 \sqrt{6} n^{3/2}}. 
\end{equation}

As in the previous section, there are subsectors of the parameter space worth discussing. First we notice that by setting $A^1 = A^2 = A^3$ we recover 5D minimal gauged supergravity, and only the R-symmetry gauge field in \eqref{eq:RFgaugefields5D} is present. On the other hand, this model also admits some subsectors with enhanced supersymmetry, which have been previously studied in\footnote{In \cite{Bomans:2024vii}, the authors imposed different boundary conditions that allow them to interpret \eqref{eq:5DU13solution} as dual to Gukow-Witten defects \cite{Gukov:2006jk}. In \cite{Couzens:2021tnv} the same local solution is given a different global completion and is studied as a disk compactification in the sector with enhanced ${\mathcal{N}}=(2,2)$ supersymmetry.} \cite{Bomans:2024vii,Couzens:2021tnv}. In particular, as discussed there, in the case $A^1 = A^2$ with $A^3=0$ the solution preserves 2D ${\cal{N}}=(2,2)$ superconformal symmmetry. Conversely, in the case $A^1 = A^2 = 0 $ with $A^3$ free, a round S$^3$ is restored in the internal space, which enhances the supersymmetry to small 2D ${\cal{N}}=(4,4)$ superconformal symmetry.

\subsection{6D Gauged Supergravity}\label{sec:6DGaugedSugra}

We conclude this Section by studying the solution dual to codimension-2 defects in the D4-brane theory. This solution, obtained at the level of the 6D U(1)$^{2}$ gauged supergravity in \cite{Boisvert:2024jrl}, reads
    \begin{equation} \label{eq:6DU12}
    \begin{split}
        \dd s^{2}_{6} & = \left(\frac{2}{g}\right)^{\frac{5}{2}}H^{\frac{1}{4}}y^{\frac{3}{2}} \rho^{\frac{1}{2}}\left( \rho^{2}\dd x^{2}_{1,2} + \frac{\dd \rho^{2}}{\rho^{2}} + \frac{\dd y^{2}}{P} + \frac{P}{H}n^{2}\dd z^{2} \right), \\[2mm]
        A^{I} & = \left(\alpha_{I} + \frac{2\kappa}{g} \sqrt{1-\frac{\nu}{q_I}}\frac{q_{I}}{h_{I}}n\right) \dd z, \qquad
        e^{2\lambda_{I}} = \frac{H^{\frac{2}{5}}y^{\frac{12}{5}}}{h_{I}}, \qquad e^{8\sigma} = \frac{g}{2 H^{\frac{1}{10}}y^{\frac{3}{5}}\rho}. \\
    \end{split}
    \end{equation}
As before $\kappa^{2}=1$, $n>0$, $y>0$ for the scalars to be well defined, and the positive functions $H$ and $P$ are given by
    \begin{equation}\label{eq:Functions6D}
        H = y^{-4}h_{1}h_{2}, \qquad
        P = H - y^{2} - \nu y^{-2},\qquad
        h_{I} = y^{4}+q_{I}, \\
    \end{equation}
where we included the non-BPS parameter $\nu$. Here, we have made a change of variables with respect to \cite{Boisvert:2024jrl}, such that setting $q_{I}=\alpha_{I}=\nu=0$, using the uplift in Appendix \ref{ap:6DGaugedSugra} and moving to the dual frame, we recover the vacuum written in the form of \eqref{eq:NewDpMetric}. 

Before proceeding to the study of defect boundary conditions, we note that this solution can be lifted to the 7D U(1)$^{2}$ AdS gauged supergravity. In 7D, this solution, which is of the form AdS$_{5}\times\Sp{1}$ foliated over an interval, has been studied as spindle-compactification \cite{Ferrero:2021wvk,Ferrero:2021etw}, disk-compactification and dual of Argyres-Douglas theories in \cite{Bah:2021hei,Bah:2021mzw}. Furthermore, it was shown in \cite{Conti:2025wwf,Gutperle:2022pgw,Gutperle:2023yrd} that it is possible to define values of the integration constants such that the solution can be interpreted as a codimension-2 monodromy defect on a 6D dual CFT. From this, it follows that it is also possible to interpret the 6D supergravity solution as a holographic dual to a 5D non-conformal theory with a codimension-2 monodromy defect. We briefly repeat the analysis in our parametrization for completeness. In the following, we choose $\kappa=1$ for convenience.

Again, we focus on the supersymmetric case, so we set $\nu=0$. The Killing spinors and supersymmetry analysis for the 6D solution, and its circle uplift in 7D, have been carried out in \cite{Ferrero:2021etw,Ferrero:2024vmz}. As shown in \cite{Ferrero:2024vmz}, for general $q_I$ the solution \eqref{eq:6DU12} preserves 3D ${\mathcal{N}}=2$ supersymmetry, we describe later a particular tuning of the parameters that leads to enhancement of supersymmetry.

We impose boundary conditions as explained in Section \ref{sec:GeneralDefect}. From \eqref{eq:SolvingqI} we find
    \begin{equation}\label{eq:qI6D}
        q_{I} = - \frac{g\, y^{4}_{\core}\mu^{I}}{2n+g\mu^{I}}.
    \end{equation}
Then, \eqref{eq:Solvingycore} leads
    \begin{equation}
         y_{\core} = \frac{1}{2n}\sqrt{ g \mu^{1}+ 2n}\sqrt{g \mu^{2}+ 2 n} .
    \end{equation}
from here we see that we must take $\mu^{I}>-2n/g$. Finally, setting $k^{2}=1$ in \eqref{eq:reg2dmetric} we find
    \begin{equation}
         g \mu^{1} + g \mu^{2}  = 1 - n.
    \end{equation}
As before, we find it convenient to parametrize the solution in terms of the monodromies of the vector fields dual to the R-symmetry and flavour currents. Since  $A_{\RR} = -A^{1}-A^{2}$ and $A_{\FF} = A^{1}-A^{2}$, we define  $\mu_{\RR} = -\mu^{1}-\mu^{2}$ and $\mu_{\FF} = \mu^{1}-\mu^{2}$. In terms of these, we have
    \begin{equation}
        y_{\core} = \frac{1}{4n}\sqrt{(1+3n)^{2}-g^{2} \mu^{2}_{\FF}}, \qquad
        g \mu_{\RR} = n- 1,
    \end{equation}
that give us also the expression of $q_I$ substituting in \eqref{eq:qI6D}.
So, as in the previous cases, a non-zero background gauge field for the R-symmetry is only possible when there is a conical deficit in the dual field theory. This was also noted in \cite{Conti:2025wwf} for the 7D uplift of the solution. Finally, using the uplift in Appendix \ref{ap:6DGaugedSugra} we find $g=1/\ell$, with $\ell$ as in Section \ref{sec:Dp-branes}. 

We now briefly illustrate two sub-sectors of the theory.

\subsubsection*{U(1) Invariant Truncation}

One case in which we can find more explicit results is taking the flavour gauge field to vanish, which is achieved by $A^1=A^2$. In this case $\mu_\FF = 0$ and 
\begin{equation}
    y_{\core} = \frac{1 + 3 n }{4 n}, \qquad q_1 = \frac{(n-1) (1 + 3n )^3}{2^8 n^4}.
\end{equation}

\subsubsection*{Supersymmetry Enhanced Sector}

Supersymmetry enhancement has been studied at the level of the 7D solution in \cite{Bah:2021mzw,Bah:2021hei,Gutperle:2023yrd,Capuozzo:2023fll}. 
From the 7D perspective, for general $q_I$ the solution preserves 4D ${\mathcal{N}}=1$ superconformal symmetry, and by setting $q_2=0$ this is increased to 4D ${\mathcal{N}}=2$. Reducing from 7D to 6D along one of the Minkowski directions inside AdS$_{5}$, breaks the AdS isometries, and the spinors charged under conformal symmetry are no longer present in 6D, which reduces the supercharges by half. Still, the analysis regarding the supersymmetry enhancement from the internal manifold works in the same way. 

In the case where one of the gauge fields is set to zero, for instance $A^{2}=0$, which can be achieved by taking $q_2 = 0$, the solution exhibits an enhancement of supersymmetry. The reason is that by setting to zero one of the gauge fields, the internal manifold restores a round S$^2$ that enhances the supersymmetry to 3D ${\mathcal{N}}=4$.  For the 3D ${\mathcal{N}}=4$ case, we have
\begin{equation}
    y = \frac{\sqrt{1 + n}}{\sqrt{2} \sqrt{n}}, \qquad q_1 = \frac{n^2-1}{4 n^2} .
\end{equation}

\section{The \texorpdfstring{$p=5$}{p=5} case} \label{sec:p5}

As stated in Section \ref{sec:Dp-branes}, the D5-brane needs to be studied separately. This is due to the fact that the coordinate change \eqref{eq:CoordChangeBraneFrame} is not well defined in this case. In fact, the D5-brane metric is not conformally $\AdS_{7}\times \Sp{3}$ but rather a linear dilaton background with metric conformal to $\mathbb{R}^{1,5}\times \mathbb{R}^{+}\times \Sp{3}$
    \begin{subequations}
    \begin{align}
        \dd s^{2}_{\text{st}} &= r\, \ell \left(\frac{\dd x^{2}_{1,5}}{\ell^{2}} + \frac{\dd r^{2}}{r^{2}} + \dd s^{2}(\Sp{3})\right), \label{eq:MetricD5Brane}\\
        F_{3} &= 2\ell^{2} \vol(\Sp{3}),\label{eq:FluxD5brane}\\
        e^{\Phi} &= \frac{r}{\ell} \label{eq:DilatonD5Brane} .
    \end{align}
    \end{subequations}
Although the dual metric does not contain an AdS factor, which is what allowed us to extend the results of \cite{Conti:2025wwf} to the $p=2,4$ branes, it is still possible extend the solution of \cite{Boisvert:2024jrl} to have a semi-infinite range for one of the spindle coordinates (see Section \ref{sec:GeneralDefect}).

We start our analysis differently: first, we consider the lift to 10D of the 7D gauged supergravity solution of \cite{Boisvert:2024jrl}, corresponding to a D5-brane wrapping a spindle. Then, as in the previous section, we change the range of the spindle coordinate $y$, which allows us to interpret the solution as a monodromy defect. We show below that the vacuum solution behaves differently when compared to the $p=2,3,4$ cases. 

The solution of \cite{Boisvert:2024jrl} is obtained in the 7D U(1)$^{2}$ gauged supergravity. In this case, the theory contains the metric, two U(1) gauge fields ($A^{I}$), and two scalars ($\lambda^{I}$), with $I=1,2$. The background configuration can be then written as
    \begin{equation}
    \begin{split}
        \dd s^2_{7} &= \frac{1}{4g^{2}}e^{\frac{2}{5}\rho}H^{\frac{1}{5}}\left(\dd x^{2}_{1,3} + \dd \rho^2 +  \frac{\dd y^2}{P} + \frac{P}{H} n^2 \dd z^2 \right), \\[2mm]
        A^{I} & = \left( \alpha_I + \frac{\kappa \, p}{2g} \sqrt{1 - \frac{\nu}{q_I}} \, \frac{y}{h_I} n \right) \dd z , \quad
        e^{\lambda_I} = e^{-\frac{1}{10}\rho}H^{\frac{1}{5}}\sqrt{\frac{p}{h_I}}\,,
    \end{split} 
    \end{equation}
where $\kappa^{2}=1$ and $H$ and $P$ are given by
    \begin{equation}
        H= h_1\,h_2 \, , \qquad 
        P=p^2\,H-y^2 - \nu y \, , \qquad
        h_I=y + q_I \, , \\
    \end{equation}
and they satisfy $H>0$ and $P>0$ for the metric to have the right signature, while the condition $p/h_{I}>0$ ensures that the scalars are real. As in the previous cases, we introduced the BPS-parameter $\nu$ that completely breaks supersymmetry. Setting $\nu=0$ leads to the supersymmetric solution of \cite{Boisvert:2024jrl}. For general $q_I$ the solution preserves four Killing spinors, the dual theory corresponds to 4D ${\mathcal{N}}=1$, where the counting is respect to the Mink$_{1,3}$. In the particular case which only one of the $q_I$ is set to zero, the supersymmetry is enhanced to 4D ${\mathcal{N}}=2$, as described in \cite{Boisvert:2024jrl}. $\alpha_I$ are constants related to the dual field theory background gauge fields $\mu^{I}$.  

At the level of the 7D solution, the solution is qualitatively different from the ones in Section \ref{sec:Defects}, since it is not conformally $\AdS_{5} \times \Sigma$, which is expected since, as explained above, the D5-brane vacuum does not have a $\AdS_{7}$ factor in dual frame\footnote{Recall that an $\AdS_{7}$ can be written as a a foliation of $\AdS_{5}\times \Sp{1}$ over an interval using \eqref{eq:ChangeToDefect}.}. 

We focus our analysis on the supersymmetric solution, so from here we take $\nu=0$. We also set $\kappa=1$ and assume $n\geq1$ for convenience. As explained in the previous sections, to obtain the defect solution we modify the range of the coordinate $y$ when compared to \cite{Boisvert:2024jrl}. As in that case, it is convenient to use the symmetry.
    \begin{equation}
        y \leftrightarrow - y, \quad
        q_{I} \leftrightarrow - q_{I},\quad
        p \leftrightarrow - p,\quad 
        z \leftrightarrow - z,\quad
    \end{equation}
to set $p>0$ and $h_{I}>0$. One difference from \cite{Boisvert:2024jrl} appears when imposing the reality condition for $P(y)$. In that case, to obtain a bounded range of the coordinate $y$, it is necessary to impose $p<1$. Instead, to have a semi-infinite range for the coordinate $y$, we impose $p>1$. Finally, this symmetry also allows us to set $y_\core\geq0$, so that we have $y\in [y_{\core},+\infty[$, which are all positive values. 

\subsection{Vacuum Solution}

Before studying defect boundary conditions, see Section \ref{sec:GeneralDefect}, let us first analyse the vacuum of this configuration, obtained by setting $q_{1}=q_{2}=\alpha_{1}=\alpha_{2}=\nu=0$. The uplift to Type IIB Supergravity, see Appendix \ref{ap:7DGaugedSugra}, reads
    \begin{subequations}
    \begin{align}
        \dd s^{2}_{\text{st}} &=  p^{-\frac{5}{2}}e^{\frac{\rho}{2}}\sqrt{y} \left( \frac{p^{2}}{4g^{2}}\left( \dd x^{2}_{1,3} + \dd \rho^{2}    + \frac{\dd y^{2}}{y^{2}(p^{2}-1)} + (p^{2}-1)n^{2}\dd z^{2}  \right) + \frac{1}{g^{2}} \dd s^{2}(\Sp{3})   \right), \label{eq:D5vacMetric} \\
        F_{3} &= \frac{2}{g^{2}}\vol(\Sp{3}), \label{eq:D5vacFlux}\\
        e^{\Phi} &= p^{-\frac{5}{2}}e^{\frac{\rho}{2}}\sqrt{y}. \label{eq:D5vacDilaton}
    \end{align}
    \end{subequations}
comparing with \eqref{eq:FluxD5brane} we identify $g=1/\ell$. This condition does not change for the defect solution. Notice that, in brane frame, the $g_{zz}$ component does not depend on $y$, we comment on this feature below. The vacuum is parametrized in a similar way to \eqref{eq:NewDpMetric}-\eqref{eq:NewDpDilaton}, where the RG flow information is encoded both in $(y,\rho)$ via \eqref{eq:ChangeToDefect}. In this case, it is also possible to move the vacuum to the canonical frame via
    \begin{equation}
        \rho = \frac{2}{p^{2}}\ln\left(\frac{r}{\ell}\right) - \sqrt{1-\frac{1}{p^{2}}} \tilde{y} + 5\ln(p), \qquad
        y = \text{exp}\left(\sqrt{1-\frac{1}{p^{2}}}\tilde{y}+ 2\left(1-\frac{1}{p^{2}}\right) \ln\left(\frac{r}{\ell}\right) \right), 
    \end{equation}
which is analogous to \eqref{eq:ChangeToDefect} for the $p=5$ case, however, the background does not take the form of \eqref{eq:MetricBraneFrame}, that is,  the canonical D$p$-brane solution with two of the spatial Minkowski directions written in polar coordinates, but rather
    \begin{subequations}
    \begin{align}
        \dd s^{2}_{\text{st}} &= r \ell \left( \frac{p^{2}}{4}\left( \dd x^{2}_{1,3} + \dd \tilde{y}^{2} + (p^{2}-1)n^{2}\dd z^{2} \right)  + \frac{\dd r^{2}}{r^{2}} + \dd s^{2}(\Sp{3}) \right),\\
        e^{\Phi} &= \frac{r}{\ell}.
    \end{align}
    \end{subequations}
Since in the solutions of our interest $z\sim z + 2\pi$, the vacuum solution is dual to 6D maximally supersymmetric SU($N$) SYM on $\mathbb{R}^{1,4}\times \Sp{1}$. We note that this does not allow to interpretation of the extension of the solution of \cite{Boisvert:2024jrl} as dual to a codimension-2 defect, since in order to do so, the vacuum should be of the form \eqref{eq:NewDpMetric}, with the $(y,z)$ space being $\mathbb{R}^{2}$ in polar coordinates, but we instead have $\mathbb{R}\times\Sp{1}$. This changes the interpretation of $n$, in this case corresponds to the size of the $\Sp{1}$.

\subsection{Boundary Conditions}

We now proceed to the study of boundary conditions for this solution. At $y=y_{\core}$ these are the same as in \ref{sec:GeneralDefect}, while for $y\rightarrow +\infty$, the condition for the gauge field remains the same, but the metric must asymptote to the vacuum studied in the section above. 

As in the previous sections, we start by \eqref{eq:SolvingqI}, from where
    \begin{equation}
      q_{I} = -y_{\core} \, (r_{I}+1), \quad
    r_{I} = -\frac{n\, p}{n\, p + 2g\, \mu^{I}}
    \end{equation}
Note that, since we are imposing $h_{I}>0$, which leads to $r_{I} y_{\core}<0$. Since we are taking $y_{\core}>0$, then $r_{I}<0$, which is equivalent to $\mu^{I}> -n\, p /(2g)$. 

We now impose regularity for the metric at $y=y_{\core}$. We start by demanding $P(y_{\core})=0$. While in the previous cases this led to a relation between $y_{\core}$ and the chemical potentials $\mu^{I}$, in this case we find
    \begin{equation}
        P(y_{\core}) = y^{2}_{\core}\left( p^{2}r_{1}r_{2}-1 \right),\quad
        H(y_{\core}) = y^{2}_{\core}r_{1}r_{2}.
    \end{equation}
One possibility is to impose $y_{\core}=0$. This however, also sets $H(y_{\core})=0$, and since $P$ and $H$ go to zero with the same power of $y_{\core}$, this leads to a finite size $\Sp{1}$ at $y=y_{\core}$, so that near $y=y_{\core}$ the $(y,z)$ space behaves as $\mathbb{R}\times \Sp{1}$ instead of $\mathbb{R}^{2}$, the later being the behaviour of our interest. In fact, one can check that setting $y_{\core}=0$ leads to the vacuum solution. Instead, to find a defect solution we must fix 
    \begin{equation}\label{eq:D5Pzero}
        p^{2}r_{1}r_{2}=1.
    \end{equation}
In this way, the parameters characterizing the solution are $p$, $n$, $y_{\core}$ and $r_{1}$, with $y_{\core}>0$ and $r_{1}<0$. Of course, one could instead invert the relation between the parameters above to $(y_{\core},r_{1}) \rightarrow \mu^{I}$, to have the later ones as free parameters, but we find this unnecessarily complicated. 

We finish the regularity analysis by requiring smoothness condition for the metric at $y=y_{\core}$, \eqref{eq:reg2dmetric}. Using \eqref{eq:D5Pzero}, we find an expression for $r_{1}$ in terms of $p$ and $n$ provided
    \begin{equation}
        p < \frac{1}{2}+ \frac{1}{2}\sqrt{\frac{4+n}{n}},
    \end{equation}
then, we find
    \begin{equation}
        r_{1} = \frac{1}{n\,p^{3}}\left( -1-n\,p \pm \sqrt{1+2n\, p- n^{2}p^{2}(p^{2}-1)} \right).
    \end{equation}
With these values, we can obtain a relation between the chemical potentials
    \begin{equation}
       g \mu^{1} + g \mu^{2} = 1.
    \end{equation}
Interestingly, this combination of the chemical potentials does not depend on $n$, as was the case in the D$p$-branes with $p=2,3,4$. 

With this, the background configuration asymptotes to a dual to 6D maximally supersymmetric SU($N$) SYM on $\mathbb{R}^{1,4}\times \Sp{1} $\eqref{eq:D5vacMetric}-\eqref{eq:D5vacDilaton} and has a smoothly shrinking $\Sp{1}$ at $y=y_{\core}$. Moreover, there are non-trivial background gauge fields for the U(1) R-symmetry and U(1) flavour symmetry. Hence, this solution can be interpreted as describing the compactification of the 6D theory on $\Sp{1}$, with background gauge fields along the $\Sp{1}$. It would be interesting to see if this solution could be connected to the ones of \cite{Nunez:2023xgl,Kumar:2024pcz}, in particular if one takes $A^{1} = A^{2}$. 

\section{Defect Entanglement Entropy}\label{sec:DefectsEE}

For backgrounds dual to codimension-2 \textit{conformal} monodromy defects in \textit{conformal} theories, it was shown in \cite{Conti:2025wwf} that, by using a direct computation of entanglement entropy using \cite{Jensen:2013lxa}, a divergence appears due to the non-compactness of the $y$ direction, for which the authors provide a renormalization procedure. 

Here, we finish our study by applying the free energy formula\footnote{This formula was proposed as a shortcut to the free energy computation via entanglement entropy.} of \cite{Macpherson:2014eza,Bea:2015fja} to the codimension-2 monodromy defect solutions of Section \ref{sec:Defects}. This requires choosing a holographic direction that realizes the holographic RG flow. For the defect backgrounds studied here, which asymptote to the vacuum \eqref{eq:NewDpMetric}-\eqref{eq:NewDpDilaton}, the RG-flow information is encoded in the coordinates $(\rho,y)$, so that a direct computation of the central charge/ free energy of the $(p+1)$-dimensional theory with a defect is not possible in this parametrization.  

By applying the formula of \cite{Macpherson:2014eza,Bea:2015fja}, treating the supergravity backgrounds as dual to a $(p-1)$-dimensional theory, we encounter the same type of divergence as in \cite{Conti:2025wwf}. We therefore extend their renormalization procedure to the cases studied here. A key element is that the backgrounds analyzed in \cite{Conti:2025wwf} take the form $\AdS_{p}\times \Sp{1}$ foliated over an interval, with an asymptotic $\AdS_{p+2}$ region at $y\rightarrow+\infty$. Although the solutions considered here, which are of the form \eqref{eq:GeneralMetric}-\eqref{eq:GeneralGaugeScalar}, do not enjoy SO($2,p+1$) conformal symmetry (except for $p=3$), the metric in dual frame has precisely the same structure as in \cite{Conti:2025wwf}. It is this fact that allows us to extend the renormalization scheme to compute the entanglement entropy of the defect theory using backgrounds dual to codimension-2 \textit{non-conformal} monodromy defects in \textit{non-conformal} maximally supersymmetric SU($N$) theories. 

With this framework in place, there is a clear procedure for the holographic computation of the codimension-2 defect free energy for $p=2,3,4$. We now describe this procedure. 

\subsection{Renormalization Scheme}\label{sec:RenormScheme}

We start by reviewing the formula of \cite{Macpherson:2014eza,Bea:2015fja}. We consider backgrounds dual to a $(d+1)$-dimensional QFT, with string frame metric of the form
    \begin{equation}\label{eq:MetricForCC}
        \dd s^{2}_{\text{st}} = a(\zeta,\theta^{i}) \left( \dd x^{2}_{1,d}+b(\zeta) \dd \zeta^{2}\right)+ g_{ij}(\zeta,\theta^i)\dd \theta^{i} \dd \theta^{j}.
    \end{equation}
Here, $\zeta$ is the holographic direction and $\theta^{i}$ are coordinates of the internal manifold. In terms of the following quantities
    \begin{equation}\label{eq:Vintdef}
        V_{\text{int}}(\zeta)= \int \dd \theta^i \sqrt{e^{-4 \Phi} a(\zeta, \theta^i)^d \det[g_{ij}]},\quad 
        H= V_{\text{int}}^2,
    \end{equation}
with $\Phi$ the 10D dilaton, the central charge, or free energy for non-conformal theories, is defined as
    \begin{equation}\label{eq:chol}
        c_{\hol} = \kappa_d d^{d} \frac{b(\zeta)^{\frac{d}{2}} H^{\frac{2d+1}{2}}}{G^{(10)}_{N} (H')^{d}},
    \end{equation}
where,  $G^{(10)}_{N} = 8\pi^{6}(\alpha')^{4} g_{s}^{2}$ is the 10-dimensional Newton constant\footnote{$\kappa_d$ is a normalization constant that we fix in each dimensions to match the conventions of \cite{Conti:2025wwf}.}. 

We apply this formula to the solutions studied in Section \ref{sec:Defects}, interpreting them as dual to a $(p-1)$-dimensional QFT, thus treating $y$ and $z$ as part of the internal manifold. Using the uplifts presented in Appendix \ref{ap:Supergravities}, for $p=2,3,4$, we obtain
    \begin{equation}\label{eq:VintCC}
        V_{\text{int}} = (\calR^{2})^{\frac{9-p}{2(5-p)}}\ell^{\frac{(7-p)^{2}}{5-p}}\rho^\frac{p-1}{5-p} \Vol(\Sp{8-p})\int^{+\infty}_{y_{\core}} \dd y\, y^\frac{p-1}{5-p},
    \end{equation}
which is clearly divergent. Here, we highlight that all the information about the defect, that is, the integration constants of the solutions, is entering only through $y_{\core}$. Moreover, recall that for the vacuum solutions $y_{\core}=1$. 

In order to obtain a finite result, we use the renormalization scheme of \cite{Jensen:2013lxa,Conti:2025wwf}. First, we introduce a UV cut-off for the integral \eqref{eq:VintCC}. We recall that, in brane frame, the metric of the defect configurations studied here asymptotes to $\AdS_{p+2}$, so it follows that it can be brought to Fefferman-Graham (FG) form by a change of coordinates $(\rho,y)\rightarrow (u,\tilde{y})$, where $u$ is the FG holographic coordinate, such that the boundary is located at $u\rightarrow+\infty$. Therefore, the UV cut-off is defined as the hypersurface $u=\LUV$. Then, inverting the change of coordinates, assuming $\rho$ is fixed and non-zero, we obtain a relation $y = y(\LUV,\rho)$ as a definition of the UV cut-off hypersurface.

As an example, let us comment on the UV regulator for the vacuum solutions \eqref{eq:NewDpMetric}. In those backgrounds, the metric in dual frame is exactly AdS$_{p+2}$ written as a foliation of $\AdS_{p}\times\Sp{1}$ over an interval. We can use \eqref{eq:ChangeToDefect} to write the metric in the form of \eqref{eq:MetricBraneFrame}. In this coordinate, the cut-off is simply defined as $u=\LUV$. Then, through \eqref{eq:ChangeToDefect}, the regulator corresponds to $y=\LUV/\rho$ in the parametrization of \eqref{eq:NewDpMetric}.

We now proceed to the definition of the UV cut-off for the defect solution. In brane frame these are of the form
    \begin{equation}\label{eq:LowerMetricForCC}
        \dd s^{2}_{p+2} = G(y) \left( \rho^{2}\dd x^{2}_{1,p-2} + \frac{\dd \rho^{2}}{\rho^{2}} + \frac{1}{P} \dd y^2 + \frac{P}{H} n^{2} \dd z^{2} \right), \\ 
    \end{equation}
where the functions $G$, $H$ and $P$ depend on the dimension but they are such that the metric is asymptotically $\AdS_{p+2}$. The coordinate transformation that brings the metric to an asymptotically Fefferman–Graham (FG) form, which is a generalization of \eqref{eq:ChangeToDefect}, was presented in \cite{Jensen:2013lxa}. For the metric \eqref{eq:LowerMetricForCC}, it takes the explicit form
    \begin{equation}\label{eq:CoordChangeFullDefect}
        u = \rho\,  \text{exp}\left(\int \dd y\, \sqrt{\frac{G-1}{P}}\right),\qquad
        \tilde{y}^{2} = \frac{1}{\rho^{2}}\,  \text{exp}\left(2 \int \dd y\, \frac{1}{\sqrt{P(G-1)}}\right) .
    \end{equation}
To define the UV regulator, it suffices to consider the asymptotic expansion of this coordinate transformation. Defining $\calF = \sqrt{\frac{G-1}{P}}$ and perform an asymptotic expansion as $y\rightarrow+\infty$, since this is the divergence we want to regulate\footnote{Also, recall that in the $(\rho,y)$ coordinates, the AdS boundary is at $y\rightarrow+\infty$. See the discussion in Section \ref{sec:Dp-branes}.}. Up to next to leading order in powers of $y^{-1}$ we have
    \begin{equation}\label{eq:ExpansioncalF}
    \calF = \frac{1}{y} +  \frac{\calCm{\alpha}}{y^{\alpha}} + \mathcal{O} \left(\frac{1}{y^{\alpha+1}}\right),
    \end{equation}
with $\alpha>1$ and it is not necessarily an integer. Here we have used the fact that, for the solutions under consideration, as $y\rightarrow +\infty$ the functions $G\rightarrow y^{2}$, $H\rightarrow y^{4}$ and $P\rightarrow y^{4}$, so that the first order in the $\calF$ expansion is always $y^{-1}$. With this, we can write the following asymptotic relation from \eqref{eq:CoordChangeFullDefect}
    \begin{equation}
        u \sim \rho\, y\, e^{ \frac{\calCm{\alpha}}{1-\alpha} y^{-\alpha+1} },  \\
    \end{equation}
which can be inverted using the Lambert $W$ function
    \begin{equation}
        y \sim \calCm{\alpha}^{\frac{1}{\alpha-1}}\, W\left(\calCm{\alpha} \left(\frac{u}{\rho}\right)^{-(\alpha-1)} \right)^{-\frac{1}{\alpha-1}}.  \\
    \end{equation}
The UV cut-off can then be found by setting $u=\LUV$ and expanding for large $\LUV$. Up to next to leading order we define the regulator $y_{\text{UV}}$ as
    \begin{equation}\label{eq:yUV}
        y_{\text{UV}} =  \frac{\LUV}{\rho}\left(1 + \frac{\calCm{\alpha}}{\alpha-1} \left(\frac{\LUV}{\rho}\right)^{1-\alpha}\right).  \\ 
    \end{equation}
We highlight that this definition of the UV cut-off is universal for all metrics of the form \eqref{eq:LowerMetricForCC} that asymptote to AdS$_{p+2}$ as $y\rightarrow + \infty$, and not only to the solutions presented here. With this, we regulate the integral in \eqref{eq:VintCC} to obtain
    \begin{equation}\label{eq:cholRegularized}
        c^{\Lambda}_{\hol} = \kappa_d \frac{2n}{\Gn{10}} \left(\frac{p-2}{p-1}\right)^{p-2} \frac{\pi^{\frac{11-p}{2}} (\calR^{2})^{\frac{1}{2}}}{\Gamma\left( \frac{9-p}{2} \right)} \left( y_{\text{UV}}^{\frac{4}{5-p}}-y_{\core}^{\frac{4}{5-p}} \right) \left( 
            \calR^{2} (\rho\ell)^{2} \right)^{-\frac{(p-3)^{2}}{2(p-5)}} \ell^{8},  \\
    \end{equation}
We note that the way in which this quantity depends on the energy scale $\rho \sim E$, that is, the factor $\left( \calR^{2}(\rho\ell)^{2} \right)^{-\frac{(p-3)^{2}}{2(p-5)}} \ell^{8}$ also appears in the $(p+1)$-dimensional free energy \eqref{eq:CCp+1} (up to $u\sim \rho$). This behaviour was already observed for the entanglement entropy of codimension-2 defect solutions, labeled ``defect entanglement entropy" (dEE) in \cite{Conti:2025wwf}, where it was found that the dEE quantity is proportional to the central charge of the theory it is embedded in. Since the computation of \cite{Conti:2025wwf} is contained within \eqref{eq:cholRegularized}, we claim  that for codimension-2 monodromy defects in $(p+1)$-dimensional maximally supersymmetric SU($N$) SYM, the dEE is proportional to the free energy of the $(p+1)$-dimensional theory. Following this observation, we write
\begin{equation}\label{eq:cLambdaCp}
    c_{\hol}^{\Lambda} = C_p \left( y_{\text{UV}}^{\frac{4}{5-p}}-y_{\core}^{\frac{4}{5-p}} \right) c^{(p)}_{\hol}, \qquad
    C_p = \kappa_d \pi n\frac{ (9-p)^p }{p^p} \frac{p-2}{p-1} {\cal{R}}, \\
\end{equation}
where we used the free energy of the $(p+1)$-dimensional theory, computed in \eqref{eq:CCp+1}.

Finally, in order to renormalize \eqref{eq:cholRegularized}, we subtract the vacuum contribution. However, this has to be weighted with a factor of $n$, to take into account that defect solutions asymptote the vacuum \eqref{eq:NewDpMetric} with $\dd z^{2}\rightarrow n^{2}\dd z^{2}$ \cite{Conti:2025wwf}. With this consideration, the renormalized quantity is given by
    \begin{equation}\label{eq:DefRenorm}
        c^{\renorm}_{\hol} = \lim_{\LUV\rightarrow +\infty} c^{\Lambda}_{\hol} - n\, c^{\Lambda,\,\text{vac}}_{\hol}. \\
    \end{equation}
Since all the information about the defect/vacuum is contained in $y_{\text{UV}}$ and $y_{\core}$, the relevant term in \eqref{eq:DefRenorm} is (recall that in the vacuum $y_{\text{UV}} = \LUV/\rho$ and $y_{\core}=1$)
    \begin{equation}
        c^{\renorm}_{\hol} 
        \propto \lim_{\LUV\rightarrow +\infty} \left[ y_{\text{UV}}^{\frac{4}{5-p}}-y_{\core}^{\frac{4}{5-p}} - \left( \left(\frac{\LUV}{\rho}\right)^{\frac{4}{5-p}} - 1 \right) \right]. \\
    \end{equation}
For the UV part, using \eqref{eq:yUV} and expanding for $\LUV\rightarrow +\infty$ we have
    \begin{equation}
    \begin{aligned}
        y_{\text{UV}}^{\frac{4}{5-p}} -  &\left(\frac{\LUV}{\rho}\right)^{\frac{4}{5-p}} \\
        &= \left(\frac{\LUV}{\rho}\right)^{\frac{4}{5-p}} \left[ 1 + \frac{4\calCm{\alpha}}{(\alpha-1)(5-p)} \left(\frac{\LUV}{\rho}\right)^{1-\alpha}  + \mathcal{O} \left( \left(\frac{\LUV}{\rho}\right)^{2-2\alpha}  \right)\right] -  \left(\frac{\LUV}{\rho}\right)^{\frac{4}{5-p}}. \\
    \end{aligned}
    \end{equation}
From here, we see that the leading order term is canceled by the renormalization scheme. In order for the next to leading order term to be finite, it is required that
    \begin{equation}\label{eq:mBound}
        \alpha \geq 1 + \frac{4}{5-p} \\
    \end{equation}
If the bound is saturated, there is a non-vanishing contribution to $c^{\renorm}_{\hol}$. We show below that, when this procedure is applied to codimension-2 defects in D$p$-brane theories ($p=2,3,4$), the bound is saturated in all cases. Finally, the dEE of the effective theory on a codimension-2 monodromy defect on $(p+1)$-dimensional maximally supersymmetric SU($N$) SYM reads
\begin{equation}\label{eq:cRenormFinal}
        c^{\renorm}_{\hol} = C_p \left( \calCm{\alpha}\, \delta_{\alpha-1,\frac{4}{5-p}} - y_{\core}^{\frac{4}{5-p}} + 1 \right) c_{\hol}^{(p)}. \\
    \end{equation}
   
We now proceed to the computation of \eqref{eq:cRenormFinal} for the codimension-2 defects in D$p$-brane theories for $p=2,3,4$.

\subsection{Supersymmetric Surface Defect in 4D SYM}\label{sec:5DefectEE}

We start with the computation of the dEE for the codimension-2 conformal defect in 4D $\mathcal{N}=4$ SU($N$) SYM, which was already computed in \cite{Conti:2025wwf}. Nevertheless, we study this case for completeness, as it allows to support the results for the non-conformal defects. 

In this case we have $G(y) = y^{\frac{2}{3}}H^{\frac{1}{3}}$, and together with \eqref{eq:5DU13functions} we find
    \begin{equation}
        \alpha = 3, \qquad
        \calCm{3} = - \frac{q_{1}+q_{2}+q_{3}}{3}.  \\
    \end{equation}
In this case ($p=3$) the bound \eqref{eq:mBound} is also saturated. Then, the defect entanglement entropy \eqref{eq:cRenormFinal} is given by\footnote{$\kappa_d=1$ in this case.}
    \begin{equation}
        c^{\renorm}_{\hol} = C_3 \left( - \frac{q_{1}+q_{2}+q_{3}}{3} - y_{\core}^{2}+1 \right) c_{\hol}^{(3)}. \\
    \end{equation}
Again, we write this expression in terms of the field theory observables, that is, the conical singularity parameter $n$ and the monodromies for the background gauge field of the flavour symmetries ($\mu_{\FF},\mu_{\FF'}$)
    \begin{equation}
        c^{\renorm}_{\hol} = \frac{\pi }{36 n} \left( 16 (n-1) (5 n+1) + 3 g^2 \left(4 \mu_{\FF}^{2}+3 \mu_{\FF'}^2\right)\right) c_{\hol}^{(3)}. \\
    \end{equation} 
Interestingly, this quantity can be expressed in terms of observables intrinsic to the defect \cite{Conti:2025wwf}: the defect Weyl anomaly $b$ and its conformal weight $h_{D}$. For backgrounds dual to supersymmetric codimension-2 conformal monodromy defects in 4D $\mathcal{N}=4$ SYM, these observables were computed in \cite{Arav:2024exg}. For our solutions 
\begin{align}
h_D & = \frac{32 (n-1) (1+ 2n)^2 + 27 g^3 \mu_{\FF'} \left(\mu_{\FF'}^2 - 4 \mu_{\FF}^2 \right) + 18 g^2 (n+1) \left(4 \mu_{\FF}^2 + 3 \mu_{\FF'}^2 \right) }{2^4 3^3 n^3} c_{\hol}^{(3)}, \\ 
b & = \frac{\pi}{2^3 3^2 n^2} \left[ 32 (n-1) \left(1 + 7n + 19 n^2\right)  +27 g^3 \mu_{\FF'} \left(\mu_{\FF'}^2 - 4 \mu_{\FF}^2 \right) + 18 g^2 (1 + 2 n) \left(4 \mu_{\FF}^2 + 3 \mu_{\FF'}^2\right)\right] c_{\hol}^{(3)}, \notag
\end{align}
from where it can be shown that the following relation holds
    \begin{equation}\label{eq:EEbhD}
        c^{\renorm}_{\hol} = \frac{1}{3} \left( b - 6 \pi n h_D \right). \\
    \end{equation}
Finally, we note that the sum of the $q_I$ is proportional to the conformal weight of the defect
\begin{equation}\label{eq:confweightqI}
    h_D = \frac{2}{3} \left( q_1 + q_2 + q_3 \right) c_{\hol}^{(3)},
\end{equation}
while the Weyl anomaly $b$ can be written as
\begin{equation}\label{eq:anomalyycore}
    b = 12 \pi n (1 - y_{\core}^2) c_{\hol}^{(3)}.
\end{equation}

\subsection{Supersymmetric 3D Defect in 5D SYM}\label{sec:6DDefectEE}

We now turn our attention to the defect solution in the D4-brane theory, obtained using the 6D gauged supergravity solution.  In brane frame, we have $G(y) = y^{\frac{6}{5}} H^{\frac{1}{5}}$, and using \eqref{eq:Functions6D} we obtain
    \begin{equation}
       \alpha = 5, \qquad \calCm{5} = - \frac{2(q_{1}+q_{2})}{5}. \\
    \end{equation}
Since in this case we have $p=4$ the bound \eqref{eq:mBound} is saturated. The defect entanglement entropy \eqref{eq:cRenormFinal} then reads
    \begin{equation}
        c^{\renorm}_{\hol} = C_{4} \left(- \frac{2(q_{1}+q_{2})}{5} - y_{\core}^{4} + 1 \right) c_{\hol}^{(4)}, \\
    \end{equation}
which, in terms of the field theory parameters ($n,\mu_{F}$), is given by 
\begin{equation}\label{eq:crenorm6Ddefect}
    c^{\renorm}_{\hol} = \frac{C_4}{2^8 5 n^4} \Big((n-1)\left(1+n\left(29+n\left(227+767 n \right) \right) \right)
+ 2(1+3n)(1+11n) g^2 \mu_{\FF}^2- g^4 \mu_{\FF}^4 \Big) c_{\hol}^{(4)}.
\end{equation}

Maximally supersymmetric Yang-Mills in dimensions different from four fails to be a conformal theory due to the gauge coupling being a dimensionful parameter. However, as explained in \cite{Kanitscheider:2008kd}, if one allows the gauge coupling to transform under conformal/Weyl transformations, as if it were a spurion, one finds a \textit{generalized conformal structure}. Therefore, it is natural to wonder if the defect entanglement entropy \eqref{eq:crenorm6Ddefect} could be decomposed in terms of quantities intrinsic to the defect as in the 4D (conformal) case, that is, if it can be written in terms of the would be conformal weight of the defect under the generalized conformal structure and the defect free energy.

In order to achieve the decomposition described above, it is convenient to recall that 5D maximally supersymmetric Yang-Mills corresponds to the 6D $\mathcal{N}=(2,0)$ superconformal theory reduced on a circle. The holographic dual of the codimension-2 monodromy defect on the 6D $\mathcal{N}=(2,0)$ theory was studied in \cite{Conti:2025wwf}, where it was showed that the defect entanglement entropy is decomposed is the same fashion as \eqref{eq:EEbhD}. Then, a circle reduction of those quantities, using the techniques of \cite{Gouteraux:2011qh}, can be used to map the 6D quantities to 5D ones. We start by noting that \eqref{eq:crenorm6Ddefect} exhibits the same dependence on the defect parameters $(n, \mu_{\FF})$ as the analogous quantity given in eq. (5.2) of \cite{Conti:2025wwf}. We elaborate more on this in the Conclusion.

\subsection{Supersymmetric Line Defect in 3D SYM}\label{sec:4DefectEE}

We finish this Section by studying the codimension-2 defect in the D2-brane theory. In this case, by moving to brane frame we find $G(y)=y^{\frac{6}{7}}H^{\frac{2}{7}}$. Then, from \eqref{eq:FunctionsD2brane} we obtain
    \begin{equation}
        \alpha = \frac{7}{3}, \quad
        \calCm{\frac{7}{3}} = - \frac{5}{14} (q_{1}+q_{2}+q_{3}).
    \end{equation}
Interestingly, although the next to leading order term in \eqref{eq:ExpansioncalF} is not an integer power of $1/y$, the bound \eqref{eq:mBound} is saturated. As stated before, this leads to an extra term in the defect entanglement entropy \eqref{eq:cRenormFinal}, which reads
    \begin{equation}
        c^{\renorm}_{\hol} = C_{2} \left( 1 - \frac{5}{14} (q_{1}+q_{2}+q_{3}) - y_{\core}^{\frac{4}{3}} \right) c_{\hol}^{(2)}.
    \end{equation}
We now write this quantity in terms of the field theory parameters: the background gauge fields of the flavour $(\mu_{\FF_{1}},\mu_{\FF_{2}})$ and R-symmetry $\mu_\RR$. For this we use \eqref{eq:ycore1st}-\eqref{eq:4dregy}, which lead to
\begin{align}
        c^{\renorm}_{\hol} & = C_{2}  \frac{1}{28 \Delta_1^{1/3} \Delta_2^{1/3}\Delta_3^{1/3} n^2} \bigg(( g\mu_{\FF_1}- g\mu_{\FF_2}-1) (2 g\mu_{\FF_1}+ g\mu_{\FF_2}+1) (g \mu_{\FF_1} + 2 g \mu_{\FF_2}-1) - 29 n^3 \label{eq:4DdEE}\\
        & - 57 n^2 + 27 n \left( (g \mu_{\FF_1})^2 + g \mu_{\FF_1} g \mu_{\FF_2}+ (g\mu_{\FF_2})^2- 1\right) + 56 n^2 \Delta_1^{1/3} \Delta_2^{1/3}\Delta_3^{1/3} \bigg) c_{\hol}^{(2)}, \notag
    \end{align}
where we used the auxiliary variables
\begin{equation}
    \Delta_1 = n + 1 - g \mu_{\FF_1} - 2 g \mu_{\FF_2}, \qquad \Delta_2 = n +1 - g \mu_{\FF_1} + g \mu_{\FF_2}, \qquad
    \Delta_3 = n + 1 + g \mu_{\FF_1} + 2 g \mu_{\FF_2}.
\end{equation}

\section{Conclusion and Future Directions}\label{sec:Conclusion}

In this paper, we computed the defect entanglement entropy for codimension-2 monodromy defect in maximally supersymmetric SU($N$) Yang-Mills theories in $(p+1)$-dimensions, ($p=2,3,4$). For $p\neq3$ neither the ambient theories nor the defects are conformally invariant,  thus generalizing the results of \cite{Conti:2025wwf,Conti:2025wyj}.

The defect geometries are obtained by considering the solutions of D$p$-branes wrapping a spindle of \cite{Boisvert:2024jrl} and changing the range of the radial coordinate of the spindle, such that now it takes values on a semi-infinite interval. Defect boundary conditions are imposed, so that at one end the geometry shrinks smoothly, while asymptotically the metric goes to the corresponding D$p$-brane one\footnote{Up to a conical singularity in the circle direction parametrized by $z$.}, but we allow the gauge fields to be asymptotically constant. This constant is related to the monodromy of background gauge fields in the dual field theory.

In order to compute the defect entanglement entropy, we employ the prescription of \cite{Macpherson:2014eza,Bea:2015fja}, interpreting the backgrounds as holographic duals to a $(p-1)$-dimensional theory. This leads to a divergence which needs to be renormalized. To address this, we exploit the fact that, in brane frame, the metric asymptotes to $\AdS_{p+2}$ written as a foliation of $\AdS_{p}\times \Sp{1}$ over an interval. We provide a renormalization prescription that allows to obtain a finite result, similar to the one of \cite{Jensen:2013lxa}. This amounts to define a UV cut-off and subtract the contribution of the vacuum. This procedure is only possible for $p=2,3,4$, since the D5-brane does not have a $\AdS_{p+2}$ asymptotic region in brane frame \cite{Kanitscheider:2008kd}.

In the conformal case ($p=3$), the renormalization procedure leads to a quantity that is proportional to a linear combination of the Weyl anomaly on the defect and its conformal weight. This feature is common to other conformal monodromy defects as showed in \cite{Conti:2025wwf,Conti:2025wyj}. In the non-conformal cases, we found that the defect entanglement entropy is proportional to the free energy of the theory where it is embedded, but it remains to be investigated if this can be written in terms of intrinsic defect quantities. 

This work provides a first step toward holographically understanding supersymmetric \textit{non-conformal} defects in \textit{non-conformal} theories, a subject which, to the best of our knowledge, remains unexplored.

\subsection{On the non-conformal dEE}\label{sec:NCdEE}

As mentioned before, maximally supersymmetric Yang-Mills shows a generalized conformal structure \cite{Kanitscheider:2008kd}, in dimensions different from three, if the gauge coupling is allowed to transform under Weyl rescalings. It is therefore natural to expect \eqref{eq:crenorm6Ddefect} and \eqref{eq:4DdEE} to be written in a form similar to \eqref{eq:EEbhD}, that is, in terms of quantities intrinsic to the defect. 

\noindent \textbf{Supersymmetric 3D Defect in 5D SYM}

As discussed previously, the connection between 5D maximally supersymmetric Yang-Mills and the 6D $\mathcal{N}=(2,0)$ theory, or equivalently, between the D4-brane in Type IIA and the M5-brane in 11D supergravity, can be used to compute 5D quantities in terms of 6D ones by means of \cite{Gouteraux:2011qh}.

Codimension-2 \textit{conformal} monodromy defects in the 6D $\mathcal{N}=(2,0)$ were studied in \cite{Conti:2025wwf}. There, it was found that the defect entanglement entropy can be written as\footnote{This formula is known to hold for a broader class of conformal 4 dimensional defects \cite{Chalabi:2021jud}. Look also at \cite{Capuozzo:2023fll} for related works.}
\begin{equation}\label{eq:EEAhD7D}
    c^{\text{renorm}}_{\text{hol}} = - (4 {\cal{A}}^{(4)} - \pi^2 n h^{(4)}_D),
\end{equation}
where ${\cal{A}}^{(4)}$ and $h^{(4)}_D$ are respectively the Weyl anomaly and the conformal weight of the defect and are explicitly given in \cite{Conti:2025wwf}. 

Reducing the background configuration along a direction parallel to the defect leads to the solution in Section \ref{sec:6DGaugedSugra}. This reduction fits in the discussion of \cite{Gouteraux:2011qh}, where it is discussed that the would be lower-dimensional anomaly is given in terms of the higher dimensional one as $\mathcal{A}^{(5)} = e^{\Phi_{0}} (2\pi R) \mathcal{A}^{(6)}$, where $R$ is the size of the S$^{1}$ in which the theory is compactified, and $e^{\Phi_{0}}$ is the asymptotic value of the lower dimensional dilaton. 

If the structure above also exists for the defect anomaly, it is natural to expect the defect entanglement entropy of the \textit{non-conformal} monodromy defect on 5D maximally supersymmetric Yang-Mills to be  
\begin{equation}\label{eq:cdefAhD}
    c_{\text{hol}}^{\text{renorm}} \sim - (4 \, \mathcal{I}^{(3)} - \pi^2 n \, h_D^{(3)}),
\end{equation} 
where $\mathcal{I}^{(3)}$ is the free energy on the 3-dimensional defect, and $h_D^{(3)}$ is the conformal weight of the defect under the generalized conformal structure of maximally supersymmetric Yang-Mills. 

Although we aim to report on this later, we note that if we identify
\begin{equation}\label{eq:AdhD6D}
\begin{split}
\mathcal{I}^{(3)} & = \kappa_d \frac{5^4 \pi}{3^2 2^{15} n^3} \left(1 + 6 n + 25 n^2 - g^2 \mu_{\FF}^2 \right) \left( 1 + 6n - 7 n^2 -g^2 \mu_{\FF}^2\right) c_{\text{hol}}^{(4)} , \\
h_D^{(3)} & = \kappa_d \frac{5^3 }{3^2 2^{11} \pi n^4} ((1+3n)^2-g^2\mu_{\FF}^2 ) (1-n(3n-2)- g^2 \mu_{\FF}^2) c_{\text{hol}}^{(4)},
\end{split}
\end{equation}
then \eqref{eq:cdefAhD} is satisfied. The reason for providing this ansatz is that the dependence of ${\cal{I}}^{(3)}$ and $h_D^{(3)}$ on the defect data $(\mu_{F},n)$ is exactly the same as the one for ${\cal{A}}^{(4)}$ and $h^{(4)}_D$ for the defect in the 6D $\mathcal{N}=(2,0)$ theory, given in \cite{Conti:2025wwf}. We expect this to be the case due to \cite{Gouteraux:2011qh}. Moreover, this is also consistent with the fact that for all codimension-2 conformal monodromy defects in maximally supersymmetric theories (\cite{Conti:2025wwf,Conti:2025wyj} and \eqref{eq:confweightqI}-\eqref{eq:anomalyycore}) the anomaly/free energy and the conformal weight of the defects can be written as
    \begin{equation}\label{eq:conjecture}
     \mathcal{A}, \mathcal{I} \, \sim \, ( 1 - y^{\chi}_{\core} ) c_{\text{hol}}^{(p)}, \qquad \qquad
     h_{D} \, \sim \, \mathcal{C}_{\alpha} c_{\text{hol}}^{(p)} \, \sim \, \sum_I q_I c_{\text{hol}}^{(p)},
    \end{equation}
where $\chi$ is a number that depends on the dimension. We expect this structure to be preserved for codimension-2 monodromy defects in maximally supersymmetric Yang-Mills.

\noindent \textbf{Supersymmetric Line Defect in 3D SYM}

A similar analysis is possible for the solution of Section \ref{sec:4DefectEE}. Since the D2-brane can be obtained as a M2-brane smeared over a circle, we can expect the line defect on the D2-brane theory of Section \ref{sec:4DefectEE} to be connected to the one on the ABJM theory, studied in \cite{Conti:2025wwf}. This relation is more subtle than the one for the M5 $\rightarrow$ D4-brane defect, since the reduction is not along one of the directions parallel to the defect, but rather along one of the internal directions. Specifically, one needs to modify the 7-sphere of the AdS$_{4}\times \Sp{7}$ solution, so that it takes the form $\Sp{6}\times \mathbb{R}$, similar to the procedure in \cite{Cvetic:1999pu}. This realises the smearing of the M2-brane, and a reduction along the $\mathbb{R}$ direction leads to the D2-brane in Type IIA. 

It is important to remark that this process does not fit into the description of \cite{Gouteraux:2011qh}: at the level of the lower dimensional gauged supergravity, both are 4-dimensional, the difference being the field content and the gauge group. Still, it would be interesting if a relation similar to \eqref{eq:EEbhD} and \eqref{eq:EEAhD7D} holds for the codimension-2 defect in the D2-brane theory, that is 
\begin{equation}\label{eq:AdhD4D}
    c_{\text{hol}}^{\text{renorm}} \sim \mathcal{I}^{(1)} - 2 \pi n h_{D}^{(1)},
\end{equation}
with $\mathcal{I}^{(1)}$ and $h_{D}^{(1)}$ the free energy and the generalized conformal weight of the defect. This relation has been proven to hold for the codimension-2 monodromy defect in ABJM theory \cite{Conti:2025wwf}, so we are encouraged to think that for the same type of defect in the D2-brane theory (which is also a maximally supersymmetric 3-dimensional theory) this relation should also hold, where the conformal symmetry of the ABJM theory is replaced by a generalized conformal structure.

Following the same structure as \eqref{eq:conjecture}, we propose
\begin{align}\label{eq:cdefIDhD}
    {\cal{I}}^{(1)} & = \kappa_d \frac{49 \pi }{24 n} \left(4 n^2- \Delta_1^{2/3}\Delta_2^{2/3}\Delta_3^{2/3} \right)c^{(2)}_{\text{hol}}, \notag \\[2mm]
    h_D^{(1)} & = \kappa_d \frac{35}{8 n^2}\frac{2}{\Delta_1^{1/3} \Delta_2^{1/3} \Delta_3^{1/3}} \big( (n+1) n^2-  ( g \mu_{\FF_1}- g \mu_{\FF_2} - 1) (2 g \mu_{\FF_1} + g \mu_{\FF_2} + 1) (g \mu_{\FF_1} + 2 g \mu_{\FF_2} - 1) \notag \\[2mm]
    & + n \left( (g \mu_{\FF_1})^2 + g \mu_{\FF_1} g \mu_{\FF_2} + (g \mu_{\FF_2})^2-1\right)\big) c^{(2)}_{\text{hol}}.
    \end{align}
We hope to report on these proposals in the future. More precisely, using the techniques of \cite{Kanitscheider:2008kd,Arav:2024exg,Arav:2024wyg}, we hope to explicitly compute these quantities and to check if \eqref{eq:cdefAhD} and \eqref{eq:AdhD4D} hold not only for monodromy defects but also for other types of defects\footnote{Since this is the case for 2D and 4D conformal defects \cite{Jensen:2018rxu,Kobayashi:2018lil,Chalabi:2021jud}.}.

\subsection{Outlook}

There are other directions we plan to address in the future.

\begin{itemize}
    \item So far, we have only dealt with supersymmetric monodromy defects. It would be interesting to check how the defect entanglement entropy or the relations proposed in Section \ref{sec:NCdEE} are modified by the absence of supersymmetry, which can be achieved by taking $\nu \neq 0$.
    \item In \cite{Kanitscheider:2008kd}, it was argued that the generalized conformal structure leads to a modification of the classical Ward identities. It is also known \cite{Arav:2024exg,Bianchi:2021snj} that conformal defects modify the Ward identities as well. It would be interesting to (holographically) study the interplay between the contribution of the non-conformal defects to the Ward identities of the generalized conformal structure.
    \item Furthermore, it would be interesting to apply the techniques of \cite{Kanitscheider:2008kd,Bobev:2025idz} to compute observables in the non-conformal cases. Moreover, precision holography computations might shed light on the proposed relations \eqref{eq:AdhD6D}-\eqref{eq:AdhD4D}, and the physical meaning of $h_{D}$. 
    \item The renormalization scheme presented here can be easily extended to the holographic duals to codimension-$n$ monodromy defects in maximally supersymmetric SU($N$) SYM, since in dual frame the metric must take the form
    \begin{equation}
        \dd s^{2}_{\dual} = G(y) \left( \rho^{2} \dd x^{2}_{1,p-n}  + \frac{\dd\rho^{2}}{\rho^{2}}  +  \frac{1}{P(y)}\dd y^{2} + \frac{P(y)}{H(y)}\dd s^{2}(\Sp{n-1})\right),
    \end{equation}
    and asymptote to $\AdS_{p+2}$ written as a foliation of $\AdS_{p+2-n}\times\Sp{n-1}$ over an interval. At the level of the $(p+2)$-dimensional gauged supergravity, these solutions contain a $(n-1)$-form potential $A_{n-1}\sim c(y)\vol(\Sp{n-1})$. This is interpreted as a background form for a ($n-2$)-form symmetry of the dual theory, which in Weyl frame has a non-trivial holonomy on $\Sp{n-1}$. Some solutions of this form have already been studied in \cite{Conti:2024qgx,Conti:2024rwd,Faedo:2025kjf,Lozano:2021fkk,Chen:2019qib,Dibitetto:2017klx,Dibitetto:2017tve,Dibitetto:2018iar,Estes:2018tnu,Capuozzo:2024onf,Apruzzi:2024ark}.
    
    The exact same prescription of Section \ref{sec:RenormScheme} works for defects of this form, which allows for a direct computation of the entanglement entropy on the effective theory on these defects. 
    \item To further study the D5-brane solution presented in Section \ref{sec:p5}. As argued above, this solution can be interpreted as dual to 6D maximally supersymmetric SU($N$) SYM on $\Sp{1}$, similar to \cite{Nunez:2023xgl,Kumar:2024pcz}. It would be interesting to understand if those two solutions are connected or if they describe two different physical systems.
    \item In \cite{Boisvert:2024jrl} a solution of 8D U(1) gauged supergravity, corresponding to a D6-brane wrapping a spindle is also considered. The procedure described in Section \ref{sec:Defects} cannot be applied to interpret this configuration as a defect. In the case of the D6-brane, the coordinate transformations \eqref{eq:CoordChangeBraneFrame} and \eqref{eq:ChangeToDefect} sets the UV region at $\rho=0$, so the boundary conditions of Section \ref{sec:Defects} do not apply here. It would be interesting to construct a solution dual to a codimension-2 defect in the D6-brane theory.  
    \item It was argued that a codimension-2 monodromy defect in maximally supersymmetric Yang-Mills on $\mathbb{R}^{1,p}$ is, after a Weyl transformation, equivalent to a twisted circle compactification of the theory on AdS$_{p}\times\Sp{1}$ (with a position dependent gauge coupling to preserve supersymmetry). It would be interesting to perform a field theory analysis similar to the one of \cite{Kumar:2024pcz} for these compactifications on AdS$_{p}\times\Sp{1}$. 
    \end{itemize}

We hope to address these and other possibly related questions in the future.

\section*{Acknowledgements}

We are thankful to Ignacio Carreño Bolla, Iñaki García Etxebarria, Jerome Gauntlett, Fridrik Gautason, Adolfo Guarino, Yolanda Lozano, Niall Macpherson, Francesco Mignosa, Carlos Nunez, Diego Rodriguez-Gomez, Christopher Rosen, Kostas Skenderis, Jesse Van Muiden, Daniel Waldram and Itamar Yaakov for useful discussion and valuable comments. The work of A.C. is supported by the Severo Ochoa fellowship PA-23-BP22-019. The work of R.S. is supported by the Ram\'on y Cajal fellowship RYC2021-033794-I. The authors acknowledge support from grants from the Spanish government MCIU-22-PID2021-123021NB-I00, MCIU-25-PID2024-161500NB-I00 and principality of Asturias SV-PA-21-AYUD/2021/52177. AC acknowledges the INGENIUM Alliance of European Universities for giving the opportunity to spend training time at the University of Crete. AC thanks Crete Center for Theoretical Physics and Imperial College London for the kind hospitality while some parts of this work were being completed.

\appendix

\section{Supergravity Conventions}\label{ap:Supergravities}

This appendix contains the Lagrangians of the different supergravities used throughout this paper. In the lower dimensional cases, we also include the uplift to Type II supergravity. 

\subsection{Type II Supergravity}\label{ap:TypeII}

We start by reviewing our conventions for Type II supergravity, for which we follow the democratic conventions of \cite{Tomasiello:2022dwe}.  The bosonic Lagrangian consists of the NS-NS sector, which contains the metric, dilaton $\Phi$ and NS 3-form $H = \dd B$, and the RR sector, which contains even(odd) forms in Type II A(B). These expressed using a polyform $F_{\pm}$ defined as
\begin{equation}
F_{\pm}= \begin{cases} 
F_0+F_2+F_4+F_6+F_8+F_{10} &\text{IIA} \\
F_1+F_3+F_5+F_7+F_9 &\text{IIB},
\end{cases}
\end{equation}
so that the upper/lower signs correspond to type IIA/IIB respectively.

In order to have the correct amount of degrees of freedom, the polyform $F$ must also obey a self duality constraint, which halves its degrees of freedom\footnote{In the democratic conventions, the Hodge dual is defined such
\begin{equation}
\star e^{\underline{M}_1...\underline{M}_k}=\frac{1}{(d-k)!}\epsilon_{\underline{M}_{k+1}...\underline{M}_{d-k}}^{~~~~~~~~~~~~~\underline{M}_1...\underline{M}_k}e^{\underline{M}_{k+1}...\underline{M}_{d-k}}, \notag
\end{equation}
where the $d=10$ indices $M$ are curved  and $\underline{M}$ are flat.}
\begin{equation}
F=\star \lambda(F),
\end{equation}
where $\lambda (C_k)= (-1)^{[\frac{k}{2}]}C_k$. 

With these considerations, the equations of motion of type II supergravity, together with the Bianchi identities for the fluxes can be written as
\begin{align}
&\dd _{H} F_{\pm}=0,~~~~ \dd H=0,~~~~ \dd (e^{-2\Phi} \star H)-\frac{1}{2}(F_{\pm},F_{\pm})_8=0, \notag \\[2mm]
& 2R - H^2-8 e^{\Phi}(\nabla)^2 e^{-\Phi}=0,~~~  R_{AB} + 2 \nabla_{A}\nabla_{B}\Phi-\frac{1}{2} H^2_{AB}-\frac{e^{\Phi}}{4} (F_{\pm})^2_{AB}=0,\label{eq:EOM}
\end{align}
where $(F_{\pm},F_{\pm})_8$ is the 8-form part of $F_{\pm}\wedge \lambda(F_{\pm})$, $\dd_{H} = \dd + H\wedge$. We have defined for a $k$-form $X$ 
\begin{equation}
(X_k)_{M}:= \iota_{\dd x^M} X_k,~~~~ X_k^2:= \sum_{k} \frac{1}{k!} (X_k)_{M_1\dots M_k}(X_k)^{M_1\dots M_k},
\end{equation}
and for a polyform $X$
    \begin{equation}
        X^2_{MN}:=\sum_{k} \frac{1}{(k-1)!} (X_k)_{MM_1\dots M_{k-1}}(X_k)_N{}^{M_1\dots M_{k-1}}.
    \end{equation}
    
Finally, the SUSY variations for the dilatino and gravitino are given by
\begin{subequations}
\begin{align}
 &\left(\nabla_M-\frac{1}{4}H_M\right) \epsilon_1 + \frac{e^\Phi}{16}F_{\pm} \Gamma_M \epsilon^2=0, \quad
 \left(\nabla_M+\frac{1}{4}H_M\right) \epsilon_2 \pm \frac{e^\Phi}{16} \lambda(F_{\pm}) \Gamma_M \epsilon^1=0, \label{eq:10dsusyeqs2}\\
&\left(\nabla-\frac{1}{4}H-\dd\Phi\right)\epsilon_1=0, \quad
 \left(\nabla+\frac{1}{4}H-\dd\Phi\right)\epsilon_2=0.\label{eq:10dsusyeqs4}
\end{align}
\end{subequations}
Here we have used the Clifford map for the polyforms. Also, the spin covariant derivative is defined as
\begin{equation}
\nabla_M= \partial_{M}+ \frac{1}{4}\omega_M{}^{\underline{P}\underline{Q}}\Gamma_{\underline{P}\underline{Q}},
\end{equation}
with $\omega_{M}^{\phantom{M}\underline{P}\underline{Q}}$ the spin connection. 

The spinors $\epsilon_{1,2}$ are Majorana-Weyl spinors. Using the chirality matrix, $\hat \Gamma =  \Gamma^{\underline{0}...\underline{9}}$, $\epsilon_{1,2}$ satisfy the chirality conditions
\begin{equation}
\hat \Gamma \epsilon_1= \epsilon_1,~~~~\hat \Gamma \epsilon_2=\mp \epsilon_2,
\end{equation}
where, as before, the upper/lower sign corresponds to Type II A/B respectively.

\subsection{7D Supergravity and its Uplift}\label{ap:7DGaugedSugra}

As explained in \cite{Boisvert:2024jrl}, we are interested in an abelian truncation - common to two models: the 7D ISO(4) and SO(4) gaugings- with two gauge fields $A^{(I)}$ and two scalars $\lambda_I$, presented in \cite{Bigazzi:2001aj}\footnote{Look at \cite{Samtleben:2005bp} for a classification of all possible gaugings in 7d maximal supergravity.}. The action of the models is
\begin{equation}\label{eq:7Daction}
S = \frac{1}{16\pi G_N^{(7)}}\int \dd ^7x \sqrt{-g}\left[R-2\mathcal{V}-5(\partial{_\mu} \lambda_+)^2-(\partial{_\mu}\lambda_-)^2  -\frac{1}{4} \sum_{I=1}^2 e^{-4\lambda_I}(F^{I})^2 \right], \\
\end{equation}
where $\lambda_{\pm}=\lambda_1\pm \lambda_2$, $F^{I}= d A^{I}$ and the scalar potential is
\begin{equation}\label{eq:potential}
\mathcal{V} = -2 g^2 e^{2(\lambda_1+\lambda_2)}. \\
\end{equation}
In this model the R-symmetry gauge field is
\begin{equation}
    A_\RR = A_1 + A_2. \\
\end{equation}
The uplift to Type IIA/IIB Supergravity has been written in \cite{Boisvert:2024jrl}. First we introduce some auxiliary quantities as
\begin{equation}
U = e^{2\lambda_1}\cos^2\eta+e^{2\lambda_2}\sin^2\eta , \qquad
V = 2 \left[ e^{4\lambda_1}\cos^2\eta + e^{4\lambda_2} \sin^2\eta- U \left(e^{2\lambda_1}+e^{2\lambda_2}\right)\right]. \\
\end{equation}
These combinations help us to write the ten dimensional metric in string frame metric as
\begin{align}
\dd s^2_{\text{st}} & = e^{2(\lambda_1+\lambda_2)} \dd s^2_7+\frac{1}{g^2}\left( \dd \eta^2+\frac{1}{U} \left(e^{2\lambda_2}\cos^2\eta (\dd  \phi_1 - g A^1)^2 + e^{2\lambda_1}\sin^2\eta (\dd  \phi_2 - g A^2)^2 \right) \right), \notag \\
B & =\frac{1}{2g^2 U}\left[e^{2\lambda_1}\cos^2\eta \, \dd \phi_1\wedge(\dd \phi_2-2g A^{2})-e^{2\lambda_2}\sin^2\eta \, (\dd \phi_1-2g A^{1})\wedge \dd \phi_2 \right], \\
e^{2\Phi} & =\frac{e^{6(\lambda_1+\lambda_2)}}{U}. \notag 
\end{align}
where $\dd s^2_7$ is the metric of the seven dimensional solution. The expression for $B$ applies in the case $A^1 \wedge A^2 = 0$, that is satisfied by the solution of interest. If this is not the case, the uplift is given in terms of its Field Strength tensor $H$, for this we refer to \cite{Boisvert:2024jrl}.

\subsection{6D Supergravity and its Uplift}\label{ap:6DGaugedSugra}

The solutions in Section \ref{sec:6DGaugedSugra} are obtained using a U(1)$^{2}$ truncation of the 6D SO(5) gauged supergravity \cite{Cowdall:1998rs}. The bosonic content of this truncation is: the metric, two gauge fields $A^{I}$, and three real scalars $\lambda_{I}$ and $\sigma$, where $I=1,2$. The bosonic Lagrangian reads \cite{Boisvert:2024jrl}
    \begin{equation}
        S = \frac{1}{16G^{(6)}_{N}} \int \dd ^{6}x \sqrt{-g}\bigg( 
        R - 2{\cal{V}} -  5(\partial \lambda_{+})^{2}- (\partial \lambda_{-})^{2} - 80(\partial\sigma)^{2} - \sum_{I=1}^2\frac{1}{2}e^{-4(\lambda_{I}+\sigma)}(F^{I})^{2}
        \bigg), \\
    \end{equation}
where $F^{I} = \dd A^{I}$, $\lambda_{\pm} = \lambda_{1} \pm \lambda_{2}$ and the scalar potential is given by
    \begin{equation}
       {\cal{V}} = -\frac{g^{2}}{4}e^{4\sigma}\left( 8 e^{2(\lambda_{1}+\lambda_{2})} + 4 e^{-4\lambda_{1}-2\lambda_{2}}
       + 4 e^{-2\lambda_{1}-4\lambda_{2}}
       - e^{-8\lambda_{1}-8\lambda_{2}}\right). \\
    \end{equation}
We introduce the R-symmetry and flavor gauge field in the same conventions as \cite{Conti:2025wwf}
\begin{equation}\label{eq:RFfields6D}
    A_\RR = - (A^1 + A^2) \qquad A_\FF = -(A^2 - A^1) \\
\end{equation}

This theory can be uplifted on $\Sp{4}$ to 10D. This uplift can be obtained in the following way\footnote{In \cite{Cvetic:2000ah} only the uplift of the metric and the dilaton is derived.}: first, the 6D theory is uplifted to 7D U(1)$^{2}$ gauged supergravity \cite{Boisvert:2024jrl} using a U(1) direction $\Sp{1}_{\varphi}$. Then, the 7D theory can be uplifted to 11D supergravity using \cite{Cvetic:1999xp}. Finally, the reducing the 11D background on $\Sp{1}_{\varphi}$ leads the following Type IIA background 
    \begin{equation}
    \begin{split}
        \dd s^{2}_{\text{st}} & = \tDelta^{\frac{1}{2}} e^{-4 \sigma} \dd s^{2}_{6} + \frac{1}{g^{2}} \tDelta^{-\frac{1}{2}} e^{-8 \sigma} \left( X_{0}^{-1} \dd \hat{\mu}_{0}^{2} + \sum_{I=1}^{2} (X^{I})^{-1} (\dd \hat{\mu}_{I}^{2} + \hat{\mu}_{I}^{2} (\dd \phi_{I} + g A^{I} )^{2}) \right),\\
        F_6 & = 2  e^{4 \sigma} g\sum_{\alpha=0}^2 \left( (X^{\alpha})^2 \hat{\mu}^{2}_{\alpha} - \tDelta X^{\alpha} \right) \vol_{6} + e^{4 \sigma} g \tDelta X^{0} \vol_{6} \\
        &\phantom{=} + \frac{1}{2g} \sum_{\alpha=0}^2 (X^{\alpha})^{-1} \star_6 \dd X^{\alpha} \wedge \dd (\hat{\mu}^{2}_{\alpha}) 
        + \frac{e^{-4 \sigma}}{2 g^2} \sum_{I=1}^2 (X^I)^{-2} \dd (\hat{\mu}_{I}^{2}) \wedge (\dd \phi_I + g A^I ) \wedge \star_6 F^I, \\
        e^{\Phi} & = \tDelta^{\frac{1}{4}} e^{- 12 \sigma}. \\
\end{split}
\end{equation}
Here $\alpha=0,1,2$. The scalars $X^{\alpha}$ are related to the 6D ones as
    \begin{equation}
        X^{I} = e^{2\lambda_{I}}, \qquad X^{0} = (X^{1}X^{2})^{-2}.
    \end{equation}
Also, $\hat{\mu}_{\alpha}$ are the embedding coordinates on S$^4$ such that $\sum_{\alpha=0}^2 \hat{\mu}^{2}_{\alpha} = 1$, and 
    \begin{equation}
        \tDelta = \sum_{\alpha=0}^2 X_{\alpha} \hat{\mu}_{\alpha}^2.
    \end{equation}
Finally, $\dd s^{2}_{6}$ is the 6-dimensional gauged supergravity metric and vol$_6$ is its volume form. 

\subsection{5D Supergravity and its Uplift}\label{ap:5DGaugedSugra}

We now present the 5D U($1$)$^3$ gauged supergravity model, following the conventions of \cite{Ferrero:2021etw}. The bosonic content of this model is, the metric, two real scalars $(\phi^{1},\phi^{2})$ and three gauge fields $A^{I}$, $I=1,2,3$. The bosonic Lagrangian reads
\begin{equation}\label{d3lagoverall}
S = \frac{1}{16 \pi G^{(5)}_{N}} \int \dd ^{5}x \sqrt{-g}\Big[R-2\mathcal{V}-\frac{1}{2} \sum_{i=1}^{2} \left(\partial \phi^i\right)^{2} 
-\frac{1}{4} \sum_{I=1}^{3}\left(X^{I}\right)^{-2}(F^{I})^{2}\Big]
-F^{1} \wedge F^{2} \wedge A^{3}, \\
\end{equation}
where $I=1,2,3$, and $F^{I}=\dd  A^{I}$. 
We defined three auxiliary scalars that satisfy the constraints $X^{I}>0$ and $X^{1} X^{2} X^{3}=1$, they can be in terms of $\phi_{1}$, $\phi_2$ appearing in \eqref{eq:5DU13solution} as
\begin{equation}
X^{1} = e^{-\frac{\phi_1}{\sqrt{6}}-\frac{\phi_2}{\sqrt{2}}}\,, \qquad
X^{2} = e^{-\frac{\phi_1}{\sqrt{6}}+\frac{\phi_2}{\sqrt{2}}}\,, \qquad
X^{3} = e^{\frac{2\phi_1}{\sqrt{6}}}\,. \\
\end{equation}
Using these auxiliary scalar fields, the potential is given by 
\begin{equation}
\mathcal{V} = - 2 g^{2}\sum_{I=1}^{3}\left(X^{I}\right)^{-1}\,. \\
\end{equation}
To define the R-symmetry and flavour symmetry gauge fields, we follow the same conventions as \cite{Arav:2024exg}
\begin{equation}\label{eq:RFgaugefields5D}
    A_\RR = A^1 + A^2 + A^3, \qquad A_\FF = A^1 - A^2, \qquad A_{\FF'} = \frac{2}{3} \left( A^1 + A^2 - 2 A^3 \right). \\
\end{equation}
The uplift of this theory to Type IIB can be found in \cite{Cvetic:1999xp}
\begin{equation}
\begin{split}\label{eq:Uplift5dTypeIIB}
\dd s^2_{\text{st}} & = \Delta^{1/2} \dd  s^2_5 + \frac{1}{g^{2}}\Delta^{-1/2}\sum_{I=1}^3 (X^I)^{-1} \left[ \dd  \hat{\mu}_I^2 + \hat{\mu}_I^2 (\dd  \phi_I + A^{I})^2 \right], \\
F_5 &  = (1 + \star_{10}) \Big\{ 2g \sum_{I=1}^3 \Big[(X^{I})^2 \hat{\mu}_I^2 - \Delta X^{I}] \text{vol}_5 + \frac{1}{2g} \sum_{I=1}^3 (X^{I})^{-1} \star_5 \dd  X^{I} \wedge \dd  (\hat{\mu}_I^2) \\
& + \frac{1}{2g^{2}} \sum_{I=1}^3 (X^{I})^{-2} \dd  (\hat{\mu}_I^2) \wedge (\dd  \phi_I + A^{I}) \wedge \star_5 F^{I} \Big\} .
\end{split}
\end{equation}
$\hat{\mu}_I$ are the embedding coordinates on S$^5$ such that $\sum_{I=1}^3 \hat{\mu}^2_I = 1$ and
\begin{equation}
    \Delta = \sum_{I =1}^3 \hat{\mu}_I^2 X^I.
\end{equation}
while $\dd s^2_5$ and vol$_5$ are correspond to the 5-dimensional gauged supergravity solution.

\subsection{4D Supergravity and its Uplift}\label{ap:4DGaugedSugra}

We now present the conventions for the solution in Section \ref{sec:4D} that can be see as the U(1)$^3$ truncation of the 4D ISO(7) gauged theory, which can be obtained as a deformation of the maximal SO(8) gauging \cite{Hull:1984yy}. The bosonic fields are the metric, three gauge fields $A^I$ and four real scalars $\phi_i$. The bosonic action is\footnote{We use conventions such that the gauge coupling $g_{\text{here}}= \frac{1}{2} g_{\text{there}}$.} \cite{Boisvert:2024jrl}
\begin{equation}
S=\frac{1}{16\pi G^{(4)}_N}\int \dd ^4 x \sqrt{-g}\left[R-2\mathcal{V}-\frac{1}{2}\sum_{i=0}^3(\partial\phi_i)^2-\frac{1}{4}\sum_{I=1}^3(X^{I})^{-2}(F^{I}_{\mu\nu})^2\right]\,,
\end{equation}
where we defined the following combinations of scalars
\begin{equation}
X^{0}=\frac{1}{2} e^{\frac{1}{2}(\phi_1+\phi_2+\phi_3)-\phi_0} \qquad
X^{1}=e^{\frac{1}{2}(\phi_1-\phi_2-\phi_3)},\qquad
X^{2}=e^{\frac{1}{2}(-\phi_1+\phi_2-\phi_3)},\qquad
X^{3}=e^{\frac{1}{2}(-\phi_1-\phi_2+\phi_3)}, \\
\end{equation}
Using these auxiliary scalar fields, the scalar potential is given by
\begin{equation}
\mathcal{V} =- 2 g^2 \left[\sum_{0\le i<j\le 3}X^{i}X^{j}-\frac{1}{2}(X^{0})^2\right] .
\end{equation}
The three gauge fields $A^I$ are associated with the U(1)$^3 \in$ ISO(7) truncation. We can define the R-symmetry and flavour symmetry gauge fields to be
\begin{equation}\label{eq:RFgaugefields4D}
    A_\RR = A^1 + A^2 + A^3, \qquad A_{\FF_1} = A^1 - A^2, \qquad A_{\FF_2} = A^2 - A^3. 
\end{equation}

The uplift was constructed in \cite{Guarino:2015qaa,Guarino:2015vca,Guarino:2017pkw}. Some details are given in \cite{Boisvert:2024jrl}.

\bibliographystyle{JHEP}

\bibliography{ref}

\end{document}